\documentclass{pasj00}

\begin{document}

\author{Mariko \textsc{Kimura}\altaffilmark{1,*},
        Keisuke \textsc{Isogai}\altaffilmark{1},
        Taichi \textsc{Kato}\altaffilmark{1},
        Kenta \textsc{Taguchi}\altaffilmark{2},
        Yasuyuki \textsc{Wakamatsu}\altaffilmark{1},
        Franz-Josef \textsc{Hambsch}\altaffilmark{3,4,5},
        Berto \textsc{Monard}\altaffilmark{6,7},
        Gordon \textsc{Myers}\altaffilmark{8},
        Shawn \textsc{Dvorak}\altaffilmark{9},
        Peter \textsc{Starr}\altaffilmark{10},
        Stephen M.~\textsc{Brincat}\altaffilmark{11},
        Enrique de \textsc{Miguel}\altaffilmark{12,13},
        Joseph \textsc{Ulowetz}\altaffilmark{14},
        Hiroshi \textsc{Itoh}\altaffilmark{15},
        Geoff \textsc{Stone}\altaffilmark{16},
        Daisaku \textsc{Nogami}\altaffilmark{1}
        }
\email{mkimura@kusastro.kyoto-u.ac.jp}

\altaffiltext{1}{Department of Astronomy, Graduate School of Science, Kyoto University, Oiwakecho, Kitashirakawa, Sakyo-ku, Kyoto 606-8502}

\altaffiltext{2}{Faculty of Science, Kyoto University, Oiwakecho, Kitashirakawa, Sakyo-ku, Kyoto 606-8502}

\altaffiltext{3}{Groupe Europe\'{e}n d'Observations Stellaires (GEOS), 23 Parc de Levesville, 28300 Bailleau l'Ev\^{e}que, France}

\altaffiltext{4}{Bundesdeutsche Arbeitsgemeinschaft f\"{u}r Ver\"{a}nderliche Sterne (BAV), Munsterdamm 90, 12169 Berlin, Germany}

\altaffiltext{5}{Vereniging Voor Sterrenkunde (VVS), Oude Bleken 12, 2400 Mol, Belgium}

\altaffiltext{6}{Bronberg Observatory, Center for Backyard Astrophysics Pretoria, PO Box 11426, 
Tiegerpoort 0056, South Africa}

\altaffiltext{7}{Kleinkaroo Observatory, Center for Backyard Astrophysics Kleinkaroo, Sint Helena 1B, PO Box 281, Calitzdorp 6660, South Africa}

\altaffiltext{8}{Center for Backyard Astrophysics San Mateo, 5 inverness Way, Hillsborough, CA 94010, USA}

\altaffiltext{9}{Rolling Hills Observatory, 1643 Nightfall Drive, Clermont, Florida 34711, USA}

\altaffiltext{10}{Warrumbungle Observatory, Tenby, 841 Timor Rd, Coonabarabran NSW 2357, Australia}

\altaffiltext{11}{Flarestar Observatory, San Gwann SGN 3160, Malta}

\altaffiltext{12}{Departamento de Ciencias Integradas, Facultad de Ciencias Experimentales, Universidad de Huelva, 21071 Huelva, Spain}

\altaffiltext{13}{Center for Backyard Astrophysics, Observatorio del CIECEM, Parque Dunar, 
Matalasca\~{n}as, 21760 Almonte, Huelva, Spain}

\altaffiltext{14}{Center for Backyard Astrophysics Illinois, Northbrook Meadow Observatory, 855 Fair Ln, Northbrook, Illinois 60062, USA}

\altaffiltext{15}{Variable Star Observers League in Japan (VSOLJ), 1001-105 Nishiterakata, Hachioji, Tokyo 192-0153}

\altaffiltext{16}{Sierra Remote Observatories, 44325 Alder Heights Road, Auberry, CA USA}

\title{ASASSN-16dt and ASASSN-16hg: Promising Candidates for a Period Bouncer}

\Received{} \Accepted{}


\SetRunningHead{Kimura et al.}{The 2016 Superoutbursts of ASASSN-16dt and ASASSN-16hg}

\maketitle

\begin{abstract}

We present optical photometry of superoutbursts in 2016 
of two WZ Sge-type dwarf novae (DNe), ASASSN-16dt and ASASSN-16hg.  
Their light curves showed a dip in brightness between the first 
plateau stage with no ordinary superhumps (or early superhumps) and 
the second plateau stage with ordinary superhumps.  
We find that the dip is produced by slow evolution 
of the 3:1 resonance tidal instability and that it would be likely 
observed in low mass-ratio objects.  
The estimated mass ratio ($q \equiv M_{2}/M_{1}$) from the period 
of developing (stage A) superhumps (0.06420(3) d) was 0.036(2) 
in ASASSN-16dt.  Additionally, its superoutburst has many properties 
similar to those in other low-$q$ WZ Sge-type DNe: long-lasting 
stage A superhumps, small superhump amplitudes, long delay of 
ordinary superhump appearance, and slow decline rate in the plateau 
stage with superhumps.  
The very small mass ratio and observational characteristics 
suggest that this system is one of the best candidates for a period 
bouncer -- a binary accounting for the missing population of 
post-period minimum cataclysmic variables.  
Although it is not clearly verified due to the lack of detection 
of stage A superhumps, ASASSN-16hg might be a possible candidate 
for a period bouncer on the basis of the morphology of its light 
curves and the small superhump amplitudes.  
\textcolor{black}{
Many outburst properties of period-bouncer candidates would 
originate from the small tidal effects by their secondary stars.  
}

\end{abstract}

\section{Introduction}

   Dwarf novae (DNe) are a subtype of cataclysmic variables 
(CVs), and are close binary systems composed of a white dwarf 
(the primary), \textcolor{black}{typically} a late-type main 
sequence star (the secondary), and an accretion disk around 
the primary.  
They go through episodic abrupt increases of luminosity which 
are called ``outbursts'' (see \cite{war95book} for a review).  

   WZ Sge-type stars are an extreme subclass of DNe, and belong 
to SU UMa-type DNe.    
They have small mass ratios, and predominantly show superoutbursts 
\textcolor{black}{
defined as long-duration (more than $\sim$2 weeks) and 
large-amplitude (more than $\sim$6 mag) outbursts with superhumps} 
(see \cite{kat15wzsge} for a review and references therein).    
The superoutbursts and superhumps are believed to be caused 
due to the tidal instability, which is triggered when the disk 
expands beyond the 3:1 resonance radius 
\citep{osa89suuma,whi88tidal,hir90SHexcess,lub91SHa,lub91SHb}.
\citet{Pdot} proposed that the superhumps are classified 
into three stages by the variations of periods and amplitudes: 
stage A superhumps with a \textcolor{black}{longer} and constant 
period and increasing amplitudes, stage B ones with 
a systematically varying period and decreasing amplitudes, 
and stage C ones with a \textcolor{black}{shorter} and constant 
period and increasing amplitudes.  
\textcolor{black}{
The most distinguishing properties of WZ Sge-type DNe are 
double-peaked modulations called ``early superhumps'' and 
rebrightenings.\textcolor{black}{\footnote{\textcolor{black}{Early 
superhumps are, however, difficult to be detected 
in high-inclination systems.  
In addition, some of WZ Sge-type DNe show no rebrightening, 
but multiple rebrightenings are exclusive to WZ Sge-type 
stars \citep{kat15wzsge}.}}} 
Early superhumps are observed at the early stage of 
the superoutburst, and have a period almost equal to 
the orbital one} \citep{kat02wzsgeESH,ish02wzsgeletter}.  
Rebrightenings are observed just after the main superoutburst 
\citep{ima06tss0222,Pdot,Pdot5}.  
The early superhumps are considered to be triggered by the tidal 
instability when the disk expands beyond the 2:1 resonance 
radius \citep{osa02wzsgehump,osa03DNoutburst}.  

   The evolutionary status of CVs which have low mass ratios, 
including WZ Sge-type DNe is still unclear (see \citet{kni11CVdonor} 
and references therein). 
One of the unsolved problems is the gap between the theoretically 
predicted and observational populations of period bouncers.  
\textcolor{black}{
Period bouncers are CVs past the period minimum, and evolve 
toward longer orbital periods due to the change of mass-radius 
relation of the secondary star.  
This change is triggered by that the thermal timescale becomes 
longer than the mass-loss timescale or that the secondary 
degenerates to a brown dwarf at the final stage of the CV 
evolution \citep{rap82CVevolution,cha09massradius}.  }
Only a few period bouncer candidates have so far been found, 
although existing theory predicts that period bouncers should 
constitute most of the CV population \citep{kol93CVpopulation}.  
\textcolor{black}{
For example, \citet{lit06j1035} and \citet{lit08eclCV} detected 
that the companion stars in 4 eclipsing CVs having periods 
close to the period minimum may be brown dwarfs by 
modeling their eclipsing light curves.  It was demonstrated 
that one of the systems has a very low-mass brown-dwarf companion 
by spectroscopic observations \citep{her16j1433}.  
\citet{und08gd522} found a CV which would have a brown-dwarf 
companion and a very low mass ratio.  \citet{avi10j1238} 
also detected a brown-dwarf binary with a likely small mass 
ratio. }

Recently, several period-bouncer candidates possibly filling 
the gap between the theories and observations have been 
discovered among WZ Sge-type DNe via photometric observations 
\citep{kat13j1222,nak14j0754j2304}.  
These objects showed peculiar rebrightenings, and have very 
small mass ratios and relatively long orbital periods as 
WZ Sge-type DNe (more than 0.06 d).  Their mass ratios were 
estimated using a new method which requires the stage A superhump 
periods and orbital periods \citep{kat13qfromstageA}.  
\textcolor{black}{
As for SSS J122221.7$-$311523, one of these candidates, 
the evidence suggesting a brown-dwarf companion has also been 
found \citep{neu17j1222}.}
\citet{nak14j0754j2304} also discussed that the detected fraction 
of these candidates can account for the theoretically expected 
population of period bouncers.  
In addition, \citet{kim16a15jd} reported that one of WZ Sge-type 
DNe, which showed a peculiar main superoutburst, may 
have a small mass ratio.  
The common properties in this kind of objects are as follows: 
(1) repeating rebrightenings or dips in brightness at the main 
superoutburst stage, (2) long-lasting stage A superhumps, 
(3) large decrease of the superhump period at the stage A 
to B transition in the objects with repeating rebrightenings, 
(4) small superhump amplitudes ($\lesssim$ 0.1 mag), 
(5) long delay of ordinary superhump appearance, 
(6) slow fading rates at the plateau stage of superoutburst 
with ordinary superhumps, and (7) large outburst amplitude 
at the time of appearance of ordinary superhumps (Table 
\ref{tab:bouncer}; Sec.~7.8 of \cite{kat15wzsge}).  

   In this paper, we report on our optical photometry of the 
2016 superoutbursts of two WZ Sge-type objects, ASASSN-16dt 
and ASASSN-16hg.  
Their outbursts were detected on April 1st, 2016 and April 30th, 
2016 by the All-Sky Automated Survey for Supernovae (ASAS-SN) 
\citep{ASASSN,dav15ASASSNCVAAS}, respectively, 
\textcolor{black}{and these two objects were regarded as bright 
CV candidates by that survey because of the large outburst 
amplitudes.\footnote{\textcolor{black}{http://www.astronomy.ohio-state.edu/asassn/transients.html}}}
ASASSN-16dt has a quiescent counterpart \textcolor{black}{PSO 
J122625.408$-$113302.953 ($g$ = \textcolor{black}{20.76(5)} mag) 
and its position is (RA:) 12h26m25.41s, (Dec:) 
-\timeform{11D33'03''} (J2000.0) \citep{fle16ps1}.}  
ASASSN-16hg has a GALEX UV source and the quiescence magnitude 
in $NUV$ band is \textcolor{black}{22.8(4)} mag.  The position 
of this object is (RA:) 22h48m41.03s, (Dec:) -\timeform{35D04'40.1''} 
(J2000.0).  
\textcolor{black}{After our observational campaigns, these two 
objects were regarded as WZ Sge-type DNe by the long delay 
of superhump appearance and/or early superhumps, and 
the rebrightenings just after the main superoutbursts.  } 
We discuss the properties of these two objects, comparing with 
those of other period-bouncer candidates.  

\begin{table*}
  \caption{Properties of candidates for a period bouncer (The candidates are limited to the DNe which have been through outbursts).}
\label{tab:bouncer}
\begin{center}
\begin{tabular}{lllccccc}
\hline
Object$^{*}$ & $P_{\rm shB}$ (d)$^{\dagger}$ & Amp$^{\ddagger}$ & Delay$^{\S}$ & Decrease$^{\#}$& Profile$^{\P}$ & Decline$^{**}$ & References$^{\dagger\dagger}$\\
\hline
MASTER J2112 & 0.060221(9) & 0.10 & $\sim$12 & 2.2\% & B & 0.127(1) & 1\\
MASTER J2037 & 0.061307(9) & 0.11 & -- & 2.2\% & B, slow & 0.052(1) & 1\\
SSS J1222 & 0.07649(1) & 0.12 & $\geq$9 & 0.93\% & E, slow & 0.020(1) & 2, 3\\
OT J1842 & 0.07234 & 0.08 & $\sim$30 & -- & E, slow & 0.045(1) & 2, 4\\
OT J1735 & -- & -- & -- & -- & slow & 0.038(1) & 5\\
OT J0754 & 0.070758(6) & 0.05 & -- & 2.0\% & slow & 0.0189(3) & 6\\
OT J2304 & 0.06635(1) & 0.13 & -- & 1.3\% & slow & 0.0340(4) & 6\\
ASASSN-14cv & 0.06045(1) & 0.07 & 14 & 2.0\% & B & 0.087(1) & 7, 8\\
PNV J1714 & 0.060084(4) & 0.09 & 11 & 2.0\% & B & 0.108(1) & 7, 8\\
OT J0600 & 0.063310(4) & 0.06 & -- & 2.1\% & B & 0.080(1) & 7, 8\\
PNV J172929 & 0.06028(2) & 0.12 & 11 & 1.7\% & D & 0.094(1) & 8\\
ASASSN-15jd & 0.064981(8) & 0.09 & 10 & -- & e & 0.088(2) & 9\\
ASASSN-15gn & 0.06364(3) & 0.10 & 11 & -- & -- & 0.0635(7) & 10\\
ASASSN-15hn & 0.06183(2) & 0.10 & 12 & 2.2\% & -- & 0.080(3) & 10\\
ASASSN-15kh & 0.06048(2) & 0.08 & $\geq$13 & 1.7\% & -- & 0.0601(6) & 10\\
ASASSN-16bu & 0.06051(7) & 0.10 & 9 & 0.62\% & slow & 0.024(1) & 10\\
ASASSN-16js & 0.06093(2) & 0.23 & 10 & 1.2\% & -- & 0.085(1) & 11\\
\hline
ASASSN-16dt & 0.064610(1) & 0.08 & $\sim$23 & 0.79\% & E, slow & 0.0282(6) & This work\\
ASASSN-16hg & 0.062371(14) & 0.12 & $\geq$6 & -- & e, B & 0.090(2) & This work\\
\hline
\multicolumn{8}{l}{\parbox{440pt}{
$^{*}$Objects' name; MASTER J2112, MASTER J2037, SSS J1222, OT J1842, OT J1735, OT J0754, OT J2304, PNV J1714, OT J0600 and PNV J172929 represent MASTER OT J211258.65$+$242145.4, MASTER OT J203749.39$+$552210.3, SSS J122221.7$-$311523, OT J184228.1$+$483742, OT J173516.9$+$154708, OT J075418.7$+$381225, OT J230425.8$+$062546, PNV J17144255$-$2943481, OT J060009.9$+$142615 and PNV J17292916$+$0054043, respectively.}} \\
\multicolumn{8}{l}{
$^{\dagger}$Period of stage B superhumps.} \\
\multicolumn{8}{l}{
$^{\ddagger}$Mean amplitude of superhumps.  Unit of mag.  } \\
\multicolumn{8}{l}{$^{\S}$Delay time of ordinary superhump appearance.  Unit of days.  } \\
\multicolumn{8}{l}{$^{\#}$Decrease rate of stage B superhump period in comparison with stage A superhump period.} \\
\multicolumn{8}{l}{
\parbox{440pt}{$^{\P}$Characteristic shapes of light curves.  B: multiple rebrightenings (type-B), D: no rebrightening (type-D), E: double superoutbursts (type-E), e: a small dip in the middle of the plateau, slow: extremely slow fading rate less than $\sim$0.05 [mag d$^{-1}$].}} \\
\multicolumn{8}{l}{$^{**}$Fading rate of plateau stage.  Unit of mag d$^{-1}$.  }\\
\multicolumn{8}{l}{
\parbox{410pt}{$^{\dagger\dagger}$1: \citet{nak13j2112j2037}, 2: \citet{kat13j1222}, 3: \citet{neu17j1222}, 4: \citet{kat13j1842}, 5: \citet{Pdot5}, 6: \citet{nak14j0754j2304}, 7: Nakata et al.~in preparation, 8: \citet{Pdot7}, 9: \citet{kim16a15jd}, 10: \citet{Pdot8}, 11: \citet{Pdot9}}} \\
\end{tabular}
\end{center}
\end{table*}

\section{Observation and Analysis}

   Time-resolved CCD photometric observations were performed 
at 11 sites by the Variable Star Network (VSNET) collaboration 
team (Table E1).  The logs of the observations of ASASSN-16dt 
and ASASSN-16hg with clear filter are given in Table E2 and E3, 
respectively.  
In this study, the data from the American Association 
of Variable Star Observers (AAVSO) 
archive\footnote{$<$http://www.aavso.org/data/download/$>$} 
are also contained.
We converted all of the observation times to barycentric 
Julian date (BJD).
We applied zero-point corrections to each observer by adding 
constants before making the analyses.  
The magnitude scale of each site was adjusted to that of 
the Berto Monard system (MLF in Table E2), where USNO-B1.0 
0784-0248445 (RA: 12h26m16.102, Dec:-\timeform{11D35'03.97''}, 
$V$ = 13.6) was used as the comparison star in the photometry of 
ASASSN-16dt, and that of the Franz-Josef Hambsch system (HaC in 
Table E2), where UCAC4 276$-$217322 (RA: 22h48m45.56s, 
Dec:-\timeform{34D58'50.8''}, $V$ = 14.4) was used as the 
comparison star in the photometry of ASASSN-16hg, respectively.  
The constancy of each comparison star was checked by nearby 
stars in the same images.  
The data reduction and the calibration of the comparison stars 
were performed by each observer.  The magnitude of each comparison 
star was measured by the AAVSO Photometric All-Sky Survey 
(APASS: \cite{APASS}) from the AAVSO Variable Star 
Database\footnote{$<$http://www.aavso.org/vsp$>$}.  

   We used the phase dispersion minimization (PDM) method 
\citep{PDM} for period analyses.  The global trends of the 
light curves were subtracted by locally-weighted polynomial 
regression (LOWESS: \cite{LOWESS}) before the PDM 
analyses.  We computed the 1$\sigma$ errors of the best 
estimated periods by these analyses using the methods of 
\citet{fer89error} and \citet{Pdot2}.  

   In estimating the robustness of the PDM result, we used 
a variety of bootstraps.  We made 100 samples, each of which 
includes randomly the 50\% of observations, and performed 
PDM analyses for the samples.  
The result of the bootstrap is represented in the form of 90\% 
confidence intervals in the resultant $\theta$ statistics.

\section{ASASSN-16dt}

\subsection{Overall Light Curve}

\begin{figure}[htb]
\begin{center}
\FigureFile(80mm, 50mm){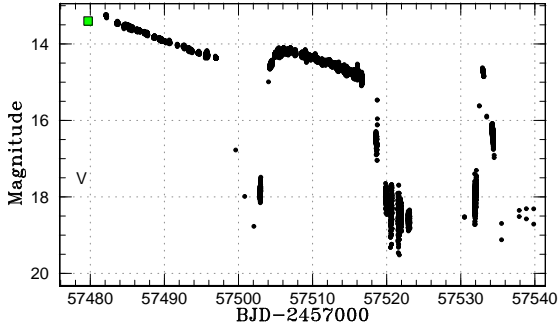}
\end{center}
\caption{Overall light curve of the 2016 superoutburst of 
ASASSN-16dt (BJD 2457478--2457540).  
The `V'-shape and quadrangle represent the upper limit and 
the detection by ASAS-SN, respectively.}
\label{overall}
\end{figure}

   We show the overall light curve of the 2016 superoutburst 
of ASASSN-16dt in figure \ref{overall}.  The superoutburst 
probably began on BJD 2457479 and the object showed a rapid 
rise at the very early stage.  
A first plateau stage continued for at least 15 days during 
BJD 2457482.1--2457497.1.  
\textcolor{black}{A dip of brightness was observed in 
the days BJD 2457499 and BJD 2457500.  }
A rapid increase in brightness 
was observed for the following $\sim$2 days, 
and the second plateau stage continued for about two weeks 
during BJD 2457504.0--2457516.8.  A rapid fading was seen on 
BJD 2457518.  
There were no observations during BJD 2457524--2457530.  
A rebrightening was detected for a few days during 
BJD 2457531.8--2457534.5.  

\subsection{Early Superhumps}

\begin{figure}[htb]
\begin{center}
\FigureFile(70mm, 50mm){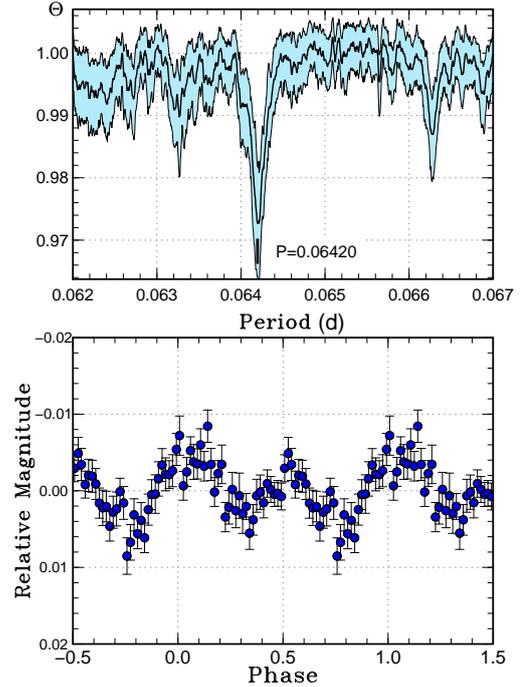}
\end{center}
\caption{Early superhumps in the 2016 superoutburst of ASASSN-16dt.  
The area of gray scale means 1$\sigma$ errors.  
Upper: $\Theta$-diagram of our PDM analysis 
(BJD 2457482.1--2457493.0).  Lower: Phase-averaged profile.}
\label{esh-a16dt}
\end{figure}

   Before the rapid decrease on BJD 2457499, double-waved 
modulations with a constant period, 0.06420(2) d, were 
detected.  
\textcolor{black}{They lasted during BJD 2457482.1--2457493.0.  
After BJD 2457483, the humps became noisy.  }  
We regard them as early superhumps.  
Figure \ref{esh-a16dt} represents the results of the PDM 
analysis and the phase-averaged profile of the early 
superhumps.  

\subsection{Ordinary Superhumps}

\begin{figure}[htb]
\begin{center}
\vspace{-1cm}
\FigureFile(80mm, 150mm){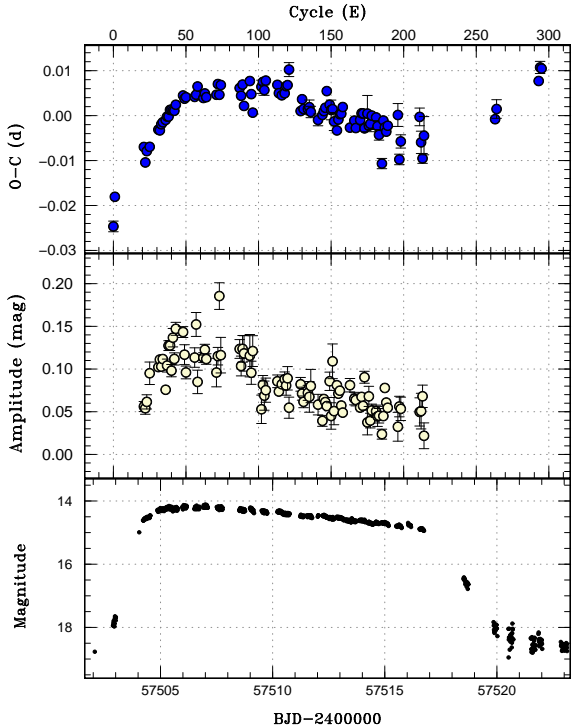}
\end{center}
\caption{Upper panel: $O - C$ curve of the times of superhump 
maxima during BJD 2457502.8--2457522.1 (the second plateau 
stage of the main superoutburst in ASASSN-16dt).  An ephemeris 
of BJD 2457502.925071$+$0.0653055 E was used for drawing this 
figure.  Middle panel: amplitudes of superhumps.  
Lower panel: light curves. The horizontal axis in units of BJD 
and cycle number is common to these three panels.}
\label{o-c}
\end{figure}

   During the dip (on BJD 2457502), ordinary superhumps 
started to develop.  The $O - C$ curve of times of superhump 
maxima, the amplitudes of superhumps, and the light curves 
during BJD 2457502.8--2457522.1 are shown in the upper panel, 
the middle panel and the lower panel of figure \ref{o-c}, 
respectively.  
We determined the times of maxima and amplitudes of ordinary 
superhumps in the same way as in \citet{Pdot}.  Some points 
with large errors were removed in calculating the $O - C$ 
and amplitudes.  
The resultant times are given in Table E4.  We regarded 
the term of stage A as BJD 2457502.8--2457506.8 ($0 \leq E 
\leq 58$) from both the $O - C$ curve and the variations of 
the superhump amplitudes.  
   We determine the term of stage B as being 
BJD 2457506.8--2457516.8 ($62 \leq E \leq 214$), from the 
nonlinear behavior on the $O - C$ curve and the decreasing 
amplitudes of superhumps.  No stage C superhumps were found.  
   The superhumps continued after the termination of the 
main superoutburst.  At the post-superoutburst stage during 
BJD 2457519.9--2457522.0 ($263 \leq E \leq 295$), the 
superhumps having a longer period than the stage B superhumps 
were detected.  
   Some modulations were seen at the rebrightening; however, 
we were not able to detect the superhump maxima and periods.  

\begin{figure*}[htb]
\begin{center}
\begin{minipage}{0.49\hsize}
\FigureFile(70mm, 50mm){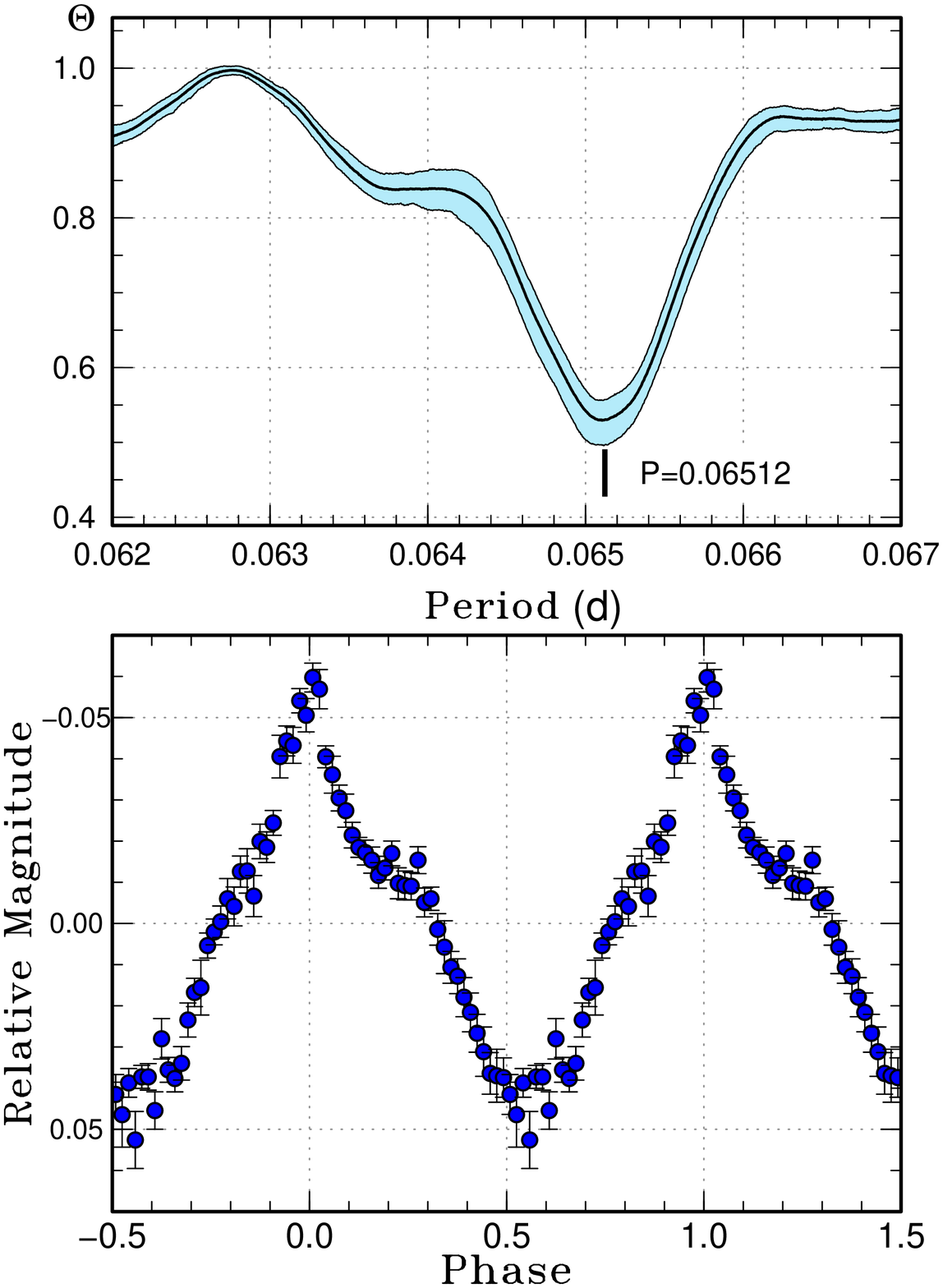}
\end{minipage}
\begin{minipage}{0.49\hsize}
\FigureFile(70mm, 50mm){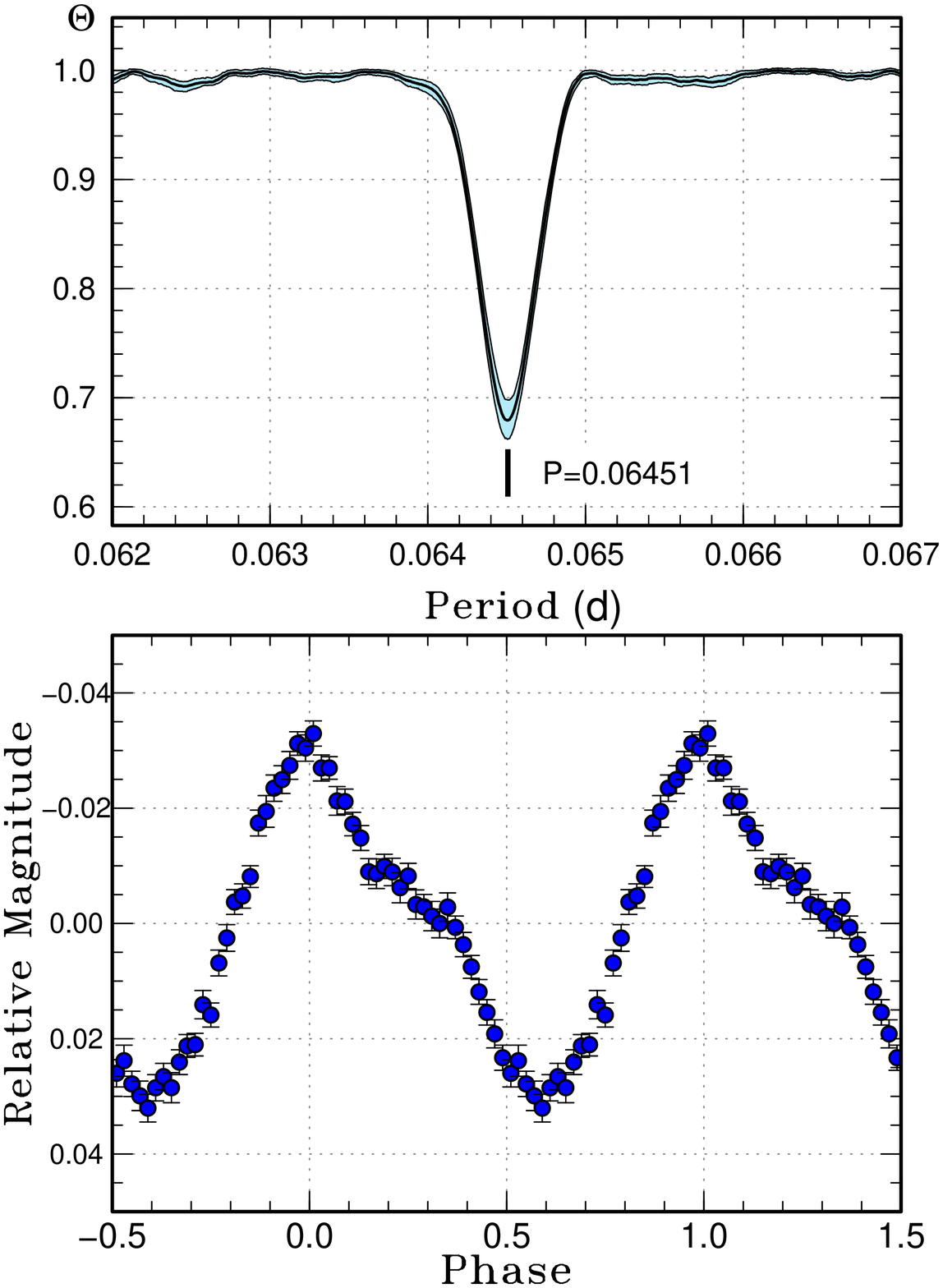}
\end{minipage}
\end{center}
\caption{Stage A and B superhumps in the second plateau stage 
of the 2016 superoutburst of ASASSN-16dt are represented in the 
left and right panels, respectively.  The area of gray 
scale means 1 $\sigma$ errors.  Upper: $\Theta$-diagrams 
of our PDM analyses.  Lower: Phase-averaged profiles.  }
\label{pdm}
\end{figure*}

   We applied period analyses by the PDM method for stage A 
and stage B, and obtained periods of $P_{\rm shA}$ = 
0.06512(1) d and $P_{\rm shB}$ = 0.064507(5) d (see 
the upper panels of figure \ref{pdm}).  
Here, the data having low accuracy were excluded from the 
light curve when we performed our PDM analyses.  
The derivative of the superhump period during stage B was 
$P_{\rm dot} (\equiv \dot{P}_{\rm sh}/P_{\rm sh}) = -1.6 (0.5) 
\times 10^{-5} {\rm s}~{\rm s}^{-1}$.  
   The mean profiles of superhumps are also shown in the 
lower panels of figure \ref{pdm}.  
   In addition, the estimated period of the superhumps 
at the post-superoutburst stage was 0.06493(4) d.  

\section{ASASSN-16hg}

\subsection{Overall Light Curve}

\begin{figure}[htb]
\begin{center}
\FigureFile(80mm, 50mm){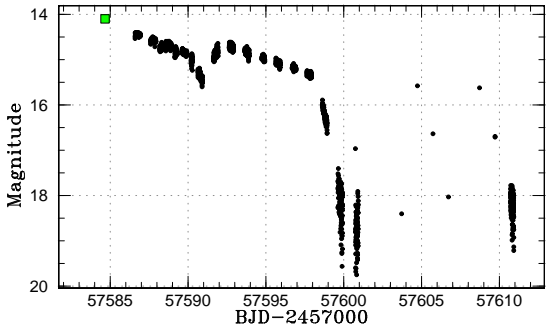}
\end{center}
\caption{Overall light curve of the 2016 superoutburst of 
ASASSN-16hg (BJD 2457584--2457611).  
The quadrangle represents the detection by ASAS-SN.}
\label{overall-16hg}
\end{figure}

   The overall light curve of the 2016 superoutburst in 
ASASSN-16hg is shown in figure \ref{overall-16hg}.  
The first plateau stage continued for more than 6 days 
before a dip in brightness on BJD 2457590.  
Soon after the dip, the system became bright again and the 
second plateau stage began and continued for a week during 
BJD 2457591.6--2457597.9.  
We detected two rebrightenings after the main superoutburst.  

\subsection{Ordinary Superhumps}

\begin{figure}[htb]
\begin{center}
\vspace{-1cm}
\FigureFile(80mm, 150mm){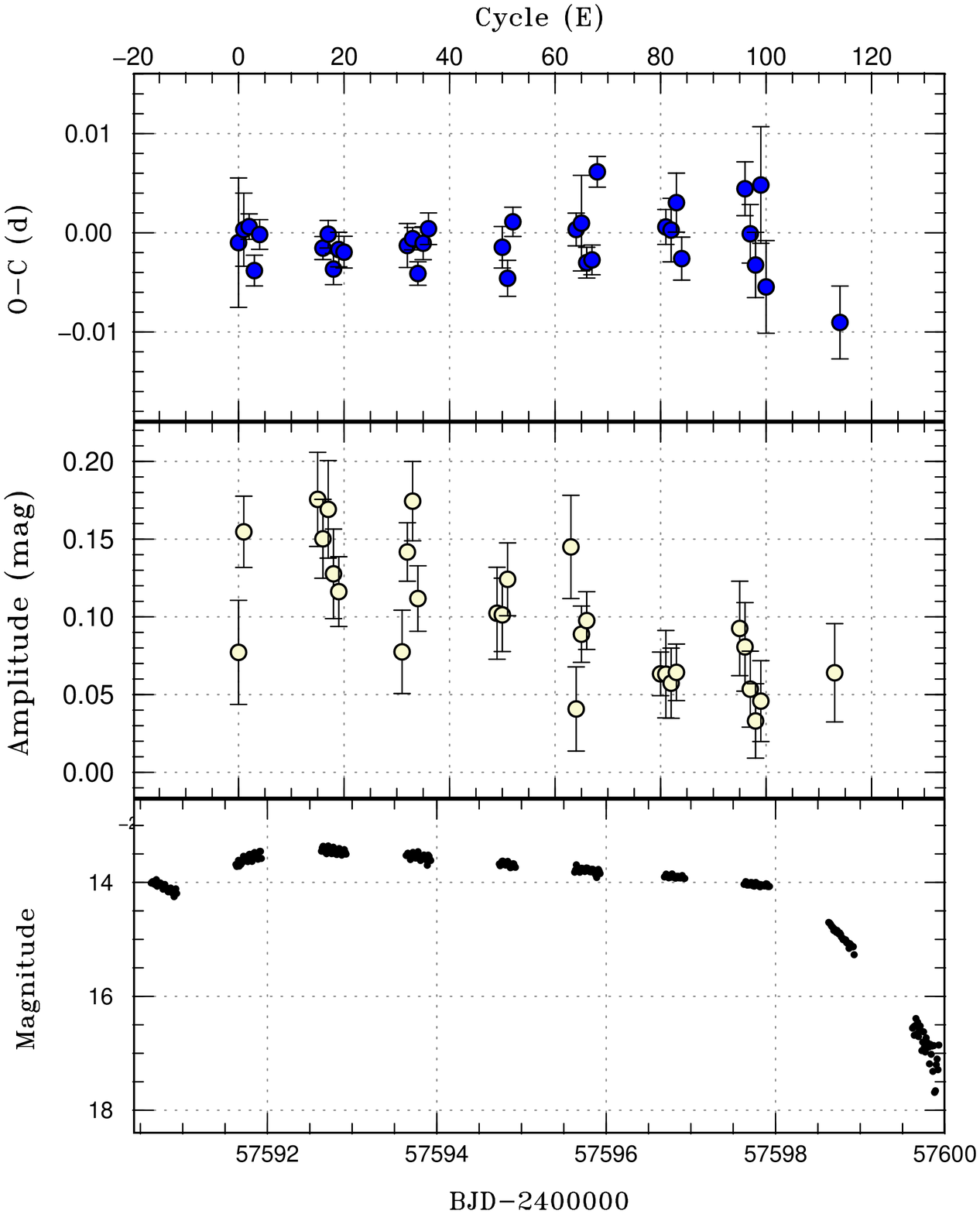}
\end{center}
\caption{Upper panel: $O - C$ curve of the times of superhump 
maxima during BJD 2457591.6--2457598.8 (the second plateau 
stage of the main superoutburst in ASASSN-16hg).  An ephemeris 
of BJD 57591.6610$+$0.0623475 E was used for drawing this 
figure.  Middle panel: amplitudes of superhumps.  
Lower panel: light curves.  The horizontal axis in units of BJD 
and cycle number is common to these three panels.}
\label{o-c-a16hg}
\end{figure}

   We have found ordinary superhumps in the second plateau 
stage and \textcolor{black}{at the beginning of the next decay}.  
   The $O - C$ curve of times of superhump maxima, the 
amplitudes of superhumps and the light curves during 
the interval are displayed in the upper, middle and 
lower panels of figure \ref{o-c-a16hg}, respectively.  
We determined the times of maxima of ordinary superhumps 
in the same way as in Sec.~3.3.  
The resultant times are given in Table E5.  We regarded the 
term during BJD 2457591.6--2457598.8 (0 $\leq$ E $\leq$ 100) 
as stage B judging from the trend of the $O - C$ curve and 
the decreasing amplitudes.  
Although there is a possibility that the stage A superhumps 
appeared between BJD 2457590 (the dip) and BJD 2457591.6 (the 
initial part of the stage B superhumps), they were not detected 
due to the low sampling rate of the data and the lack of 
observations.  
   We were not able to determine whether some modulations before 
the dip on BJD 2457590 were early superhumps or not since their 
small amplitudes and the sparse data prevented us from confirming 
double-waved variations.  
\textcolor{black}{We note that the different superoutburst 
stages were not clearly distinguished in the $O-C$ curve.  }

\begin{figure}[htb]
\begin{center}
\FigureFile(70mm, 50mm){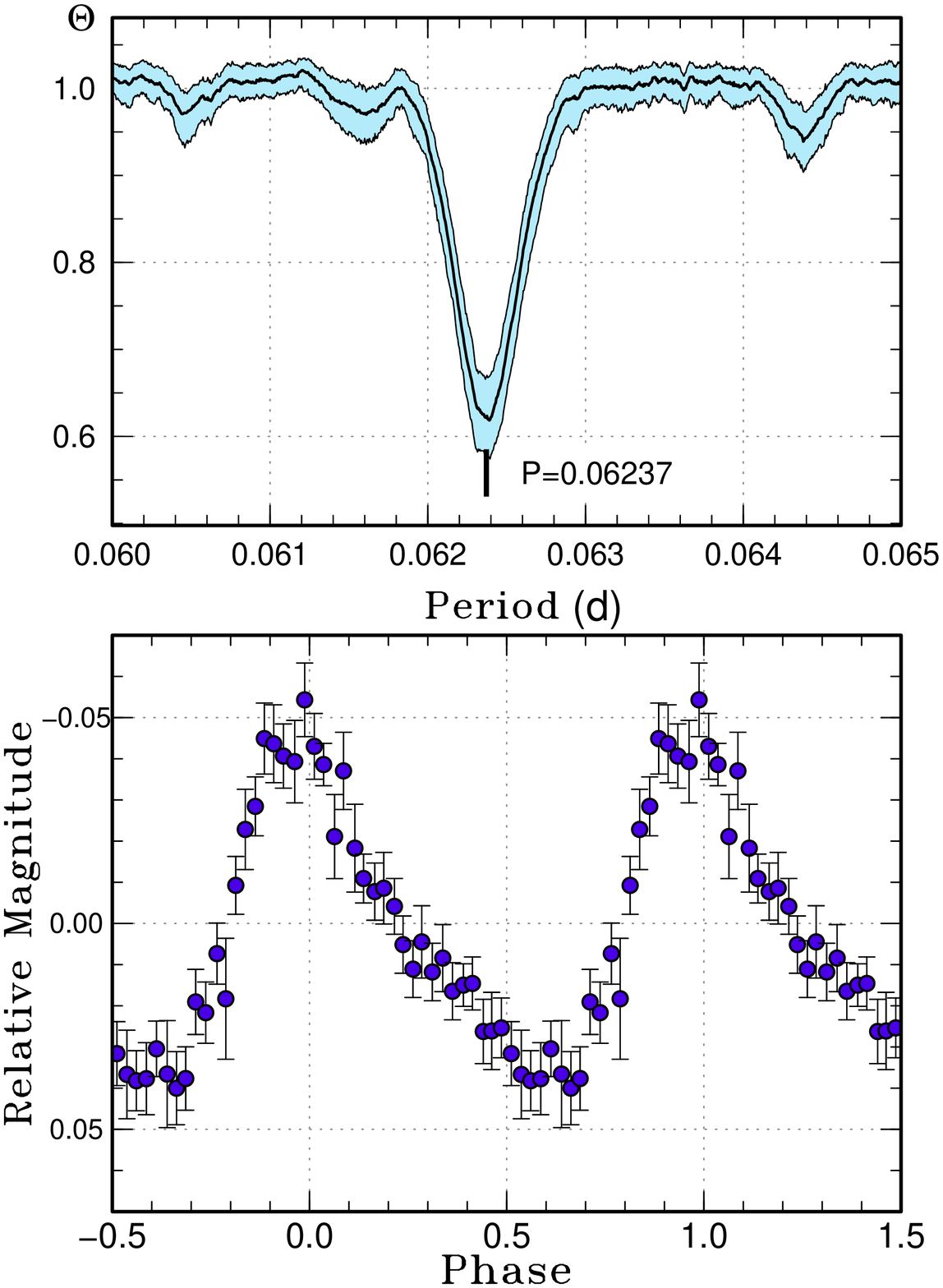}
\end{center}
\caption{Stage B superhumps in the second plateau stage 
of the 2016 superoutburst of ASASSN-16hg are represented.  
The area of gray scale means 1 $\sigma$ errors.  
Upper: $\Theta$-diagram of our PDM analysis.  Lower: 
Phase-averaged profile.  }
\label{pdm-16hg}
\end{figure}

   We applied a period analysis by the PDM method for stage B 
and obtained the period of $P_{\rm shB}$ = 0.06237(1) d 
(see the upper panel of figure \ref{pdm-16hg}).  
The derivative of the superhump period during stage B was 
$P_{\rm dot} = 0.6 (1.7) \times 10^{-5} {\rm s}~{\rm s}^{-1}$.  
   The mean profile of the stage B superhumps is also shown 
in the lower panel of figure \ref{pdm-16hg}.  

\section{Discussion}

\subsection{Mass Ratio Estimation from Stage A Superhumps}

   We can estimate the mass ratio in ASASSN-16dt using 
the method proposed by \citet{kat13qfromstageA}, assuming 
that the early superhump period is identical to the orbital 
period \citep{kat02wzsgeESH,ish02wzsgeletter}.  
\textcolor{black}{
According to \citet{hir90SHexcess}, the dynamical precession 
rate, $\omega_{\rm dyn}$ is expressed as follows: 
\begin{equation}
\omega_{\rm dyn} / \omega_{\rm orb} = Q(q)R(r), 
\label{dynamicalprecession}
\end{equation}
where the $Q(q)$ and $R(r)$ are the functions of a mass ratio 
and a given radius in an accretion disk, respectively (see 
equations (1) and (2) in \citet{kat13qfromstageA} for 
the detailed expressions).  
Under the assumption that the stage A superhumps are 
the representation of the dynamical precession at 
the 3:1 resonance radius, we can derive the value of 
mass ratio by substituting the 3:1 resonance radius, 
which is expressed as a function of 
the mass ratio as $r_{3:1} = 3^{-2/3} (1+q)^{-1/3}$, 
into equation (\ref{dynamicalprecession}).}

\begin{figure}[htb]
\begin{center}
\FigureFile(80mm, 100mm){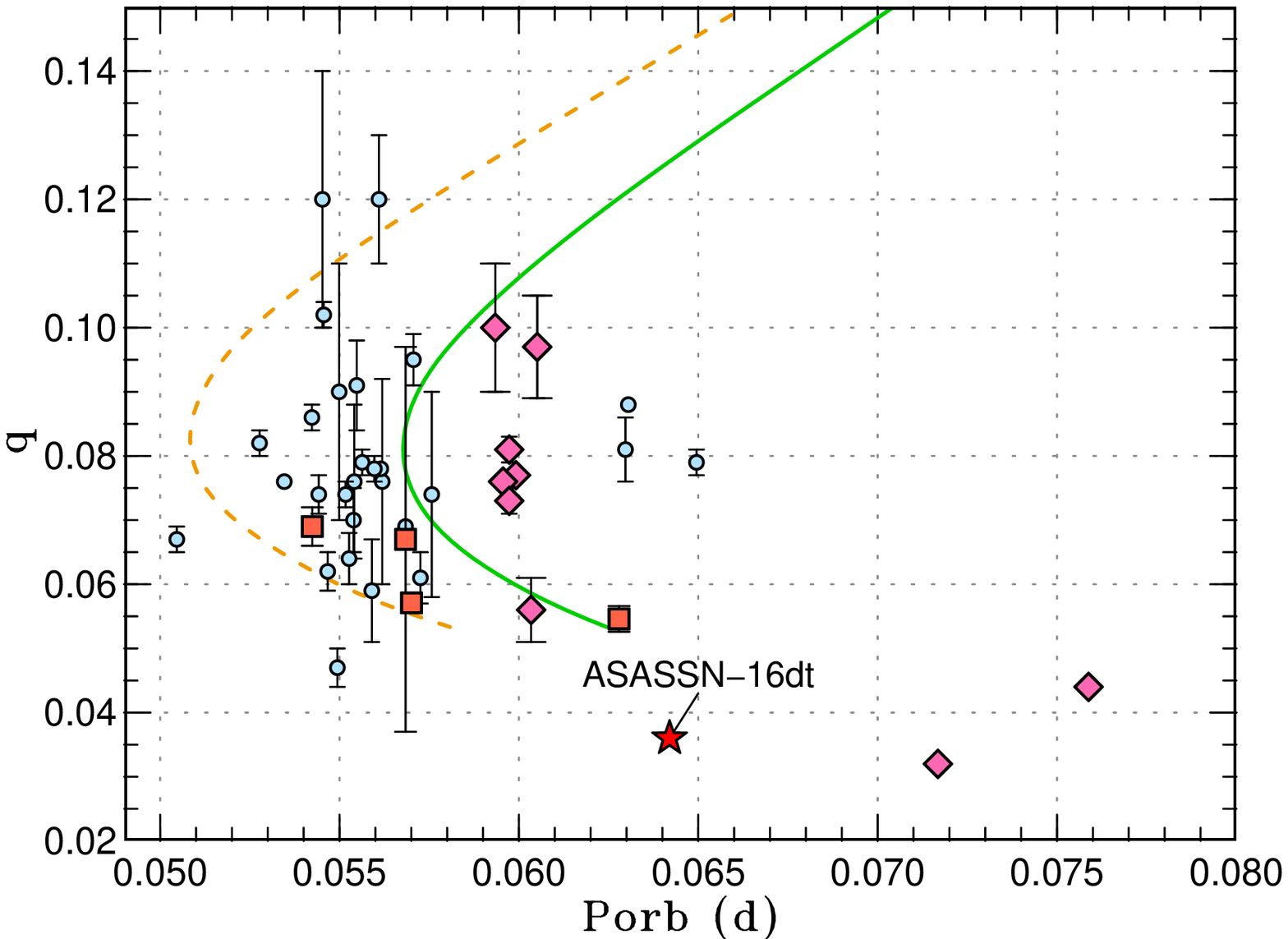}
\end{center}
\caption{$q-P_{\rm orb}$ relation of the candidates for a period bouncer \textcolor{black}{and ordinary WZ Sge-type DNe}.  The star, diamonds, rectangles and circles represent ASASSN-16dt, other candidates for a period bouncer among the 
identified WZ Sge-type DNe, the candidates for a period bouncer among eclipsing CVs, and ordinary WZ Sge-type DNe.  \textcolor{black}{The dash and solid lines represent an evolutionary track of the standard evolutional theory and that of the modified evolutional theory, respectively, which are derived from \citet{kni11CVdonor}.}}
\label{massratio}
\end{figure}

The estimated mass ratio with this method is $q$ = 0.036(2) 
as for ASASSN-16dt. This is shown on 
the $q - P_{\rm orb}$ plane in figure \ref{massratio} with 
the mass ratios of other period-bouncer candidates and 
ordinary SU UMa-type DNe derived from \citet{Pdot9}.  
\textcolor{black}{The derived errors originate from 
the errors of period estimations in Sec.~3.  }
The very small mass ratio suggests ASASSN-16dt is one of 
the best candidates for a period bouncer.  
\textcolor{black}{Our result agrees with the empirical 
relation between rebrightening types and mass ratios, 
which was suggested in \citet{kat15wzsge}.  
The $q$ values of candidates for a period bouncer are 
displayed in Table \ref{qvalues} with those of ordinary 
WZ Sge-type dwarf novae.  
Several objects close to the period minimum, which showed 
repeating rebrightenings, have been shown to be promising 
period-bouncer candidates because they share the outburst 
properties with the extremely low-$q$ period-bouncer 
candidates, and have longer orbital periods than the group of 
WZ Sge-type DNe whose orbital periods are close to the 
theoretical period minimum (see also Table \ref{tab:bouncer}).  
The three objects having longer orbital periods than 
the promising period-bouncer candidates in figure 
\textcolor{black}{\ref{massratio}} 
are not regarded as being of this kind, since they do not 
share the aforementioned properties (see also the 4th 
paragraph in Sec.~1).  
}

\begin{table*}
  \caption{Mass ratios of candidates for a period bouncer.}
\label{qvalues}
\begin{center}
\begin{tabular}{llllc}
\hline
Object$^{*}$ & $P_{\rm orb}$ (d)$^{\dagger}$ & $P_{\rm shA}$ (d)$^{\ddagger}$ & $q$$^{\S}$ & References$^{\parallel}$\\
\hline
SDSS J1507 & 0.046258 & -- & 0.0625(4) & 1\\
SDSS J1433 & 0.054241 & -- & 0.069(3) & 2, 3\\
SDSS J1501 & 0.056841 & -- & 0.067(3) & 1\\
SDSS J1035 & 0.057007 & -- & 0.057(1) & 4\\
ASASSN-16bu & 0.0593(1) & 0.06089(7) & 0.10(1)$^{\star}$ & 5\\
PNV J1714 & 0.059558(3) & 0.06130(2) & 0.076(1)$^{\star}$ & 6, 7\\
PNV J172929 & 0.05973 & 0.06133(7) & 0.073(2)$^{\star}$ & 7\\
MASTER J2112 & 0.059732(3) & 0.06158(5) & 0.081(2)$^{\star}$ & 8\\
ASASSN-14cv & 0.059917(4) & 0.06168(2) & 0.077(1)$^{\star}$ & 6, 7\\
ASASSN-16js & 0.060337(5) & 0.0617(1) & 0.056(5)$^{\star}$ & 9\\
MASTER J2037 & 0.0605(2) & 0.0627(1) & 0.097(8)$^{\star}$ & 8\\
SDSS J1057 & 0.062792 & -- & 0.055(2) & 1\\
ASASSN-16dt & 0.06420(3) & 0.06512(1) & 0.036(2)$^{\star}$ & This work\\
OT J1842 & 0.07168(1) & 0.07287(8) & 0.042(3)$^{\star}$ & 10\\
SSS J1222 & 0.07625(5) & 0.07721(1) & 0.032(2)$^{\star}$ & 11, 12\\
\hline
\multicolumn{5}{l}{\parbox{310pt}{
$^{*}$Objects' name; MASTER J2112, MASTER J2037, SSS J1222, OT J1842, PNV J1714, PNV J172929, SDSS J1057, SDSS J1035, SDSS J1433, SDSS J1501, SDSS 1433, and SDSS 1507 represent MASTER OT J211258.65$+$242145.4, MASTER OT J203749.39$+$552210.3, SSS J122221.7$-$311523, OT J184228.1$+$483742, PNV J17144255$-$2943481, PNV J17292916$+$0054043, SDSS J105754.25$+$275947.5, SDSS J103533.02$+$055158.3, SDSS J143317.78$+$101123.3, SDSS J150137.22$+$550123.4, SDSS J143317.78$+$101123.37, and SDSS J150722.30$+$523039.8, respectively.}} \\
\multicolumn{5}{l}{
$^{\dagger}$Orbital period.  } \\
\multicolumn{5}{l}{
$^{\ddagger}$Period of stage A superhumps.} \\
\multicolumn{5}{l}{
$^{\S}$\parbox{310pt}{Mass ratio.  The index $\star$ represents the mass ratio derived by the method in \citet{kat13qfromstageA}.  }} \\
\multicolumn{5}{l}{
\parbox{310pt}{$^{\parallel}$1: \citet{mca17j1057}, 2: \citet{lit08eclCV}, 3: \citet{her16j1433}, 4: \citet{sav11CVeclmass}, 5: \citet{Pdot8}, 6: Nakata et al.~in preparation, 7: \citet{Pdot7}, 8: \citet{nak13j2112j2037}, 9: \citet{Pdot9}, 10: \citet{kat13qfromstageA}, 11: \citet{kat13j1222}, 12: \citet{neu17j1222}.}} \\
\end{tabular}
\end{center}
\end{table*}

\textcolor{black}{
It is shown that some of the objects given 
in Table \ref{qvalues} would have brown-dwarf 
secondaries 
\citep{lit08eclCV,sav11CVeclmass,her16j1433,neu17j1222}.  
There may be a possibility that these objects 
come from zero-age detached white-dwarf 
and brown-dwarf binaries.  Actually, one object which 
seems to be a pre-CV candidate having a brown-dwarf 
secondary has recently found \citep{rar17wd1202}.
According to \citet{pol04CVwithBD}, however, $\sim$80 \% 
of this kind of detached binaries have small orbital 
periods less than 0.054 d, and the fraction of them 
to the total population of zero-age CVs seems to be 
smaller than a half.  
Thus some objects in Table \ref{qvalues}, which we 
discuss here, would be formed from the zero-age CVs 
with main-sequence companions via the mass loss of 
the companions rather than the zero-age CVs with 
brown-dwarf companions.  
This kind of discussion was also stated in 
\citet{neu17j1222}.}

\subsection{Dip in Brightness during Main Superoutburst}

   The dips in brightness during the main superoutbursts 
were observed in both ASASSN-16dt and ASASSN-16hg.  
\textcolor{black}{Rebrightenings in WZ Sge-type stars are 
classified into five types according to the profiles of 
the light curves: type-A (long duration rebrightening), 
type-B (multiple rebrightening), type-C (single 
rebrightening), type-D (no rebrightening), and type-E 
(double superoutbursts consisting of a plateau stage 
with early superhumps and another plateau stage with 
ordinary superhumps) \citep{ima06tss0222,Pdot,Pdot5}, }
and ASASSN-16dt belongs to WZ Sge-type objects with 
double superoutbursts.  This object is an analogue of 
SSS J122221.7$-$311523 and OT J184228.1$+$483742 
\citep{kat13j1222,kat13j1842,neu17j1222}.  
ASASSN-16hg is the second object showing an intermediate 
light curve between the single plateau stage (in type-A--D 
rebrightenings) and the double ones (in type-E rebrightening).  
The details of the classification with the morphology of 
plateau stages are described in \citet{kim16a15jd}.  
In addition, the duration of stage A superhumps is normally 
long in the candidates \citep{kat13j1222,nak14j0754j2304}, 
and also in ASASSN-16dt, the stage A superhumps continued 
for a long interval, $\sim$4 days.  

   The characteristic morphology of the light curves 
seems to represent the slow development of the 3:1 resonance 
which is believed to cause ordinary superhumps since the 
resonance keeps the disk in the hot state in WZ Sge-type DNe 
after the disappearance of the 2:1 resonance \citep{osa03DNoutburst}.  
   \citet{lub91SHa} proposed that the growth time of the 3:1 
resonance tidal instability is inversely proportional to the 
square of the mass ratio.  The small mass ratio and the 
morphology of the 2016 superoutburst in ASASSN-16dt are in 
good agreement with this theory.  
In addition, the slow growth of the 3:1 resonance is expected 
to produce the long-lasting stage A superhumps in small-$q$ 
systems.  
   Although the small dip during the main superoutburst in 
ASASSN-16hg may suggest that this system likely has a small 
$q$ value, it is unclear whether the ordinary superhumps 
slowly developed, due to the lack of detection of the initial 
part of the developing ordinary superhumps (stage A superhumps) 
(see Sec.~4.2).

\subsection{Long Delay of Superhump Appearance}

\begin{figure}[htb]
\begin{center}
\FigureFile(80mm, 50mm){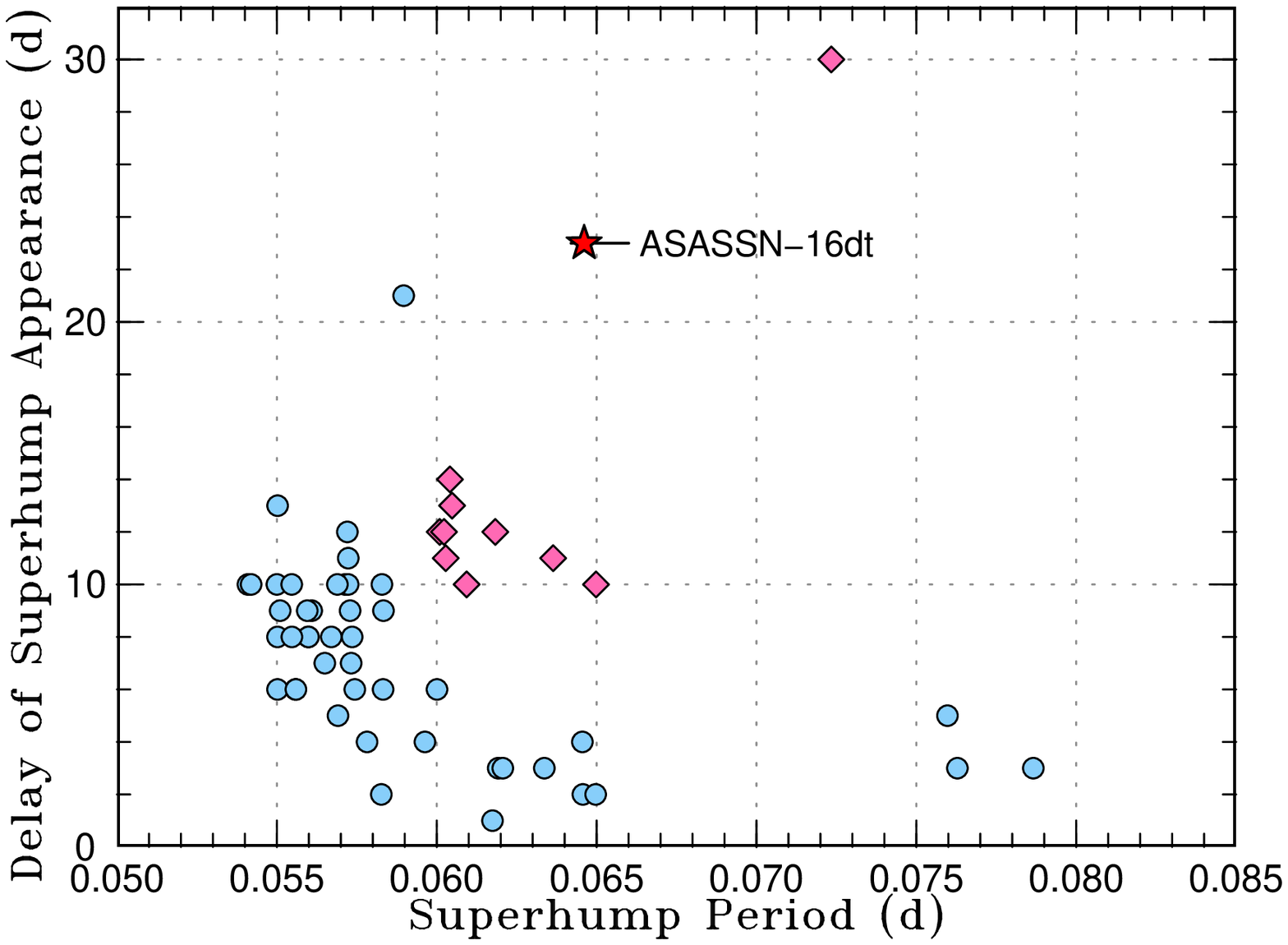}
\end{center}
\caption{$P_{\rm SH}$ vs.~delay time of ordinary superhump 
appearance.  The circles and diamonds indicate ordinary 
WZ Sge-type stars derived from Fig.~19 in \citet{kat15wzsge} 
and the candidates for a period bouncer.  The star represents 
ASASSN-16dt.  }
\label{delay}
\end{figure}

   The delays of ordinary superhump appearance are typically 
long in the candidates for a period bouncer (see Table 
\ref{tab:bouncer}), whilst those in ordinary WZ Sge-type 
stars are concentrated between 5--10 [d] \citep{kat15wzsge}.  
This feature is also clearly confirmed in the 2016 superoutburst 
of ASASSN-16dt.  
   Figure \ref{delay} shows the relation between superhump 
period and delay time in WZ Sge-type DNe.  
   On the other hand, it is uncertain whether the delay of 
the superhump appearance was long in ASASSN-16hg since there was 
no observation between the upper limit by ASAS-SN on BJD 
2457573 and the detection by the same survey on BJD 2457584.  

   In ASASSN-16dt, early superhumps were clearly detected 
at the early stage of its superoutburst.  
\citet{osa03DNoutburst} proposed that the 2:1 resonance 
causing early superhumps suppresses the 3:1 resonance.  
The presence of early superhumps lasting for at least 
$\sim$10 days in ASASSN-16dt support this explanation.  
\textcolor{black}{
As for the objects having very small mass ratios, 
the disk radius would expand far beyond the 2:1 resonance 
radius when an outburst is triggered.  This is because 
the stored disk mass naturally become large due to 
extremely low viscosity in the quiescent disk, 
while the tidal effect by the secondary is weak in 
these objects \citep{osa02wzsgehump}.  }
Collectively, these effects sustain the 2:1 resonance 
for a long interval.  

\subsection{Small Amplitude of Superhumps}

\begin{figure}[htb]
\begin{center}
\FigureFile(80mm, 50mm){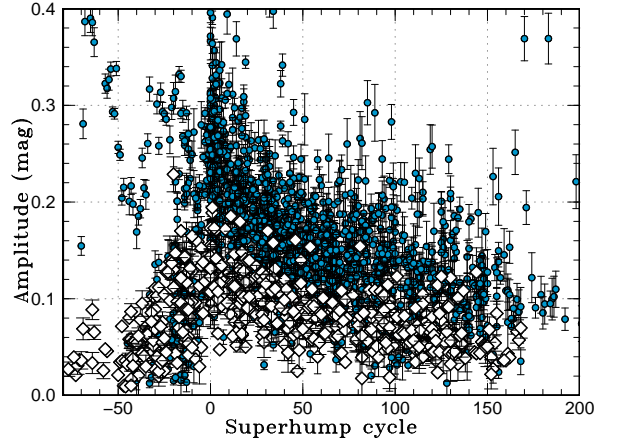}
\end{center}
\caption{Variation of superhump amplitudes in the SU UMa-type objects with 0.06 d $<$ $P_{\rm orb}$ $\leq$ 0.07 d.  
The diamonds and circles represent the candidates for a period bouncer and ordinary SU UMa-type DNe, respectively.  The data of the period-bouncer candidates are derived from \citet{nak13j2112j2037,kat13j1222,nak14j0754j2304,Pdot7,Pdot8,kim16a15jd,Pdot9}, and those of ordinary SU UMa-type DNe are derived from  \citet{Pdot,Pdot2,Pdot3,Pdot4,Pdot5,Pdot6,Pdot7,Pdot8,Pdot9}.  }
\label{amplitude}
\end{figure}

   \citet{kim16a15jd} pointed out that the average superhump 
amplitudes are small, less than 0.1 mag, in most of the 
candidates for a period bouncer (Table \ref{tab:bouncer}).  
Our results on ASASSN-16dt and ASASSN-16hg reinforce this 
observation (see also figures \ref{o-c} and \ref{o-c-a16hg}).  
   We compare the variation of superhump amplitudes of the 
candidates for a period bouncer and those of ordinary 
SU UMa-type systems having orbital periods ranging 
between 0.06--0.07 d in figure \ref{amplitude} as in 
\citet{kim16a15jd}.\footnote{We 
excluded the data with large errors more than 0.03 mag and 
of eclipsing systems.  }
In plotting this figure, we measured the amplitudes using the 
template fitting method described in \citet{Pdot} and took 
the starting point of the cycle count from the start of stage B.  
\textcolor{black}{
Since superhump amplitudes are known to depend on the 
orbital periods and the inclination angles \citep{Pdot3}, 
the data of ASASSN-16js are excluded from this figure.  
This object may have high inclination, which is judged from 
the large amplitudes of its early superhumps, $\sim$0.2 mag 
\citep{Pdot9}.  It is known that the higher the inclination is, 
the larger the amplitudes of early superhumps are 
\citep{kat15wzsge}, and the $\sim$0.2-mag amplitudes of early 
superhumps are comparable to those of WZ Sge having the 
inclination of 77$\pm$2 deg \citep{ste07wzsge}.  }
   The median value of the amplitudes between $-3 < E < 5$ 
in the period-bouncer candidates is significantly smaller, 
0.074 mag, than that in ordinary SU UMa-type DNe, 0.22 mag.  

\textcolor{black}{
   During the 3:1 resonance, the disk becomes elliptical 
due to the tidal force of the secondary; the orbiting 
secondary passes the major axis of the disk with 
the superhump period when the ordinary superhumps are 
observed.
Some particles in the tail of the eccentric disk are 
periodically absorbed to the secondary, and the time 
variations of the released energy by the viscous dissipation 
in the outer disk corresponds the superhump variations 
\citep{hir90SHexcess}.  
It has been proposed that the tidal torques exerted by the secondary 
significantly affect viscous dissipation in the outer disk 
\citep{ich94tidal}.  
   In the small-$q$ objects, the tidal force from the secondary 
would be less than that in the large-$q$ objects.  
The disk would be less elliptical and the liberated energy 
by the tidal dissipation would be small.  
Thus the reason why period-bouncer candidates show small-amplitude 
superhumps seems to be related to the weak tidal effect by 
the secondary in small-$q$ objects.
}

\subsection{Slow Fading Rate of Plateau Stage}

\begin{figure}[htb]
\begin{center}
\FigureFile(80mm, 50mm){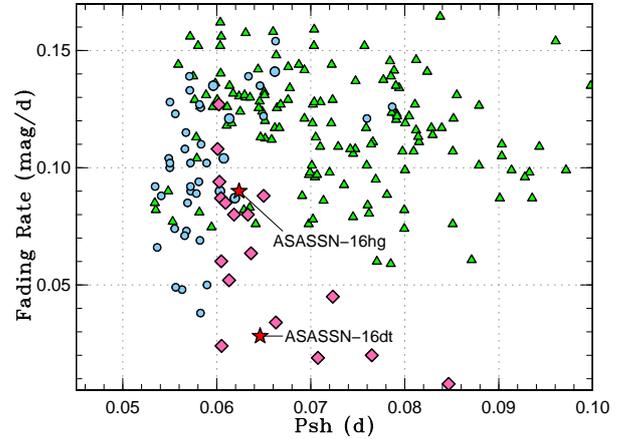}
\end{center}
\caption{Fading rate vs.~superhump period in stage B.  The circles, triangles, and diamonds represent ordinary SU UMa-type DNe, WZ Sge-type DNe, and candidates for the period bouncer, respectively. The stars indicate ASASSN-16dt and ASASSN-16hg.  The data of the ordinary SU UMa-type DNe and WZ Sge-type DNe are derived from \citet{Pdot5}.  }
\label{fading}
\end{figure}

\textcolor{black}{
   The fading rates of the plateau stage with ordinary 
superhumps in the superoutbursts of period-bouncer candidates 
are often small (Sec.~7.8 in \cite{kat15wzsge}).  
In particular, all of the three period-bouncer candidates 
including ASASSN-16dt, which showed double superoutbursts, 
had extremely low decline rates (see Table 
\ref{tab:bouncer}).  They also have very small mass ratios, 
and the durations of the superoutbursts are long -- 
more than 40 days (see also \cite{kat13j1222}).  
One period-bouncer candidate with type-B rebrightenings 
and four other candidates whose rebrightening types have 
not yet been identified showed slow declines (see also Table 
\ref{tab:bouncer}).  
The relation between the fading rate and the superhump period 
is shown in figure \ref{fading}.  
The median value of the fading rates in the period-bouncer 
candidates is 0.06 mag d$^{-1}$, while that in ordinary 
WZ Sge-type stars is 0.10 mag d$^{-1}$.  
}

\textcolor{black}{
   The decline timescale is proportional to $\alpha^{-0.7}$ 
(equation (49) in \citet{osa89suuma}).  
Here $\alpha$ represents the viscous parameter in the hot 
state, and the combination of ordinary $\alpha$ due to 
magnetohydrodynamical instability plus the viscosity 
resulting from the tidal torque 
\citep{BalbusHawley,ich94tidal}.  The slow fading rate 
in period bouncers would therefore be attributed to
the weaker tidal torque in low-$q$ objects.  
}

\section{Conclusions}

   We have reported on our photometric observations of two 
WZ Sge-type DNe, ASASSN-16dt and ASASSN-16hg, and discussed 
their similar properties to those of period-bouncer 
candidates.  The important findings are summarized as follows: 
\begin{itemize}
\item ASASSN-16dt and ASASSN-16hg underwent outbursts with a dip 
in brightness at their main superoutbursts.  This implies that 
the 3:1 resonance grew slowly in their outbursts, and that 
these objects have low mass ratios.  
\item 
\textcolor{black}{
The mass ratio in ASASSN-16dt, estimated from the method 
of \citet{kat13qfromstageA} via the stage A superhump period 
(0.06512(1) d) is 0.036(2), which is much lower than 
the theoretically expected mass ratio at the period minimum.  
The relatively long orbital period estimated from the early 
superhumps and the very low mass ratio are enough to judge 
that this object is one of the best period-bouncer candidates.  
This object also showed many features similar to those 
in other candidates for a period bouncer, featuring long-lasting 
stage A superhumps and early superhumps, small-amplitude 
superhumps, and a slow decline rate at the plateau stage.  
}
\item Although it is uncertain whether the development of the 
superhumps in ASASSN-16hg was slow due to no detection of 
the stage A superhumps, this object might be a possible 
period-bouncer candidate on the basis of the morphology of 
the plateau stage which resembles that during the 2015 
superoutburst in ASASSN-15jd \citep{kim16a15jd} and 
its small superhump amplitude.  
\item
Many outburst properties of the period-bouncer candidates would 
be explained by the small tidal effect by the secondary 
in small-$q$ systems.  
\end{itemize}

   The outburst behavior of candidates for a period bouncer 
is different from that of ordinary WZ Sge-type stars.  
\textcolor{black}{
It should be confirmed whether this behavior is inherent to 
the period bouncers, by identifying observational properties 
of many candidates.  
}
   Some of the candidates given in Table \ref{qvalues} have 
not experienced outbursts.  It would be interesting to monitor 
their behavior when they enter outbursts.

\section*{Acknowledgements}

This work was financially supported by the Grant-in-Aid for 
JSPS Fellows for young researchers (MK, KI) and by a Grant-in-Aid 
``Initiative for High-Dimensional Data-Driven Science through 
Deepening of Sparse Modeling'' from the Ministry of Education, 
Culture, Sports, Science and Technology (MEXT) of Japan 
(25120007, TK). 
Also, it was partially supported by the RFBR grant 15-02-06178.  
We appreciate All-Sky Automated Survey for Supernovae (ASAS-SN) 
detecting a large amount of DNe.  
We are thankful to many amateur observers for providing a lot of 
data used in this research.
We are grateful to Ra'ad Mahmoud for correcting the English 
in this paper.  
We appreciate an anonymous referee's giving us helpful comments.  

\section*{Supporting information}

Additional supporting information can be found in the online version 
of this article:
Supplementary tables E1, E2, E3, E4 and E5.

\newcommand{\noop}[1]{}


\renewcommand{\thetable}{%
  E\arabic{table}}
  \setcounter{table}{1}

\begin{table*}[htb]
  \caption{List of Instruments.  }
\label{tab:listtel}
\begin{center}
\begin{tabular}{lccc}
\hline
CODE$^{*}$ & Telescope (\& CCD) & Observatory (or Observer) & Site \\
\hline
BSM & 25cmSC+Moravian G2-1600 & Flarestar Observatory & San Gwann, Malta \\
deM & 35cmSC+QSI-516wsg & Observatorio Astronomico del CIECEM & Huelva, Spain \\
DKS & 25cmACF & Rolling Hills Observatory & USA \\
HaC & 40cmIRDK+FLI-ML16803 & Remote Observatory Atacama Desert (ROAD) & San Pedro de Atacama, Chile \\
Ioh & 30cmSC+ST-9XE CCD & Hiroshi Itoh & Tokyo, Japan \\
MLF & 30cmRCX400+ST8-XME & Berto Monard Calitzdorp & South Africa \\
& 35cmRCX400+ST8-XME & & \\
MGW & 43.1cmPlanewaveCDK & Gordon Myers & Siding Spring, Australia\\
& +FLI PL4710 CCD & & \\
KU1 & 40cmSC+Apogee U6 & Kyoto U.~Team & Kyoto, Japan \\
SGE & 43cmCDK+STXL-11002 & Sierra Remote Observatories & Auberry, CA, USA \\
SPE & 51cmPlanewaveCDK & Warrumbungle \& Dubbo Observatory & Australia \\
& +SBig STL 168303 & & \\
UJH & 23cmSCT+QSI-583ws & Joseph Ulowetz & USA \\
\hline
\multicolumn{4}{l}{
$^{*}$see the annotation in Table E2.  } \\
\end{tabular}
\end{center}
\end{table*}

\begin{table*}[htb]
\caption{Log of observations of the 2016 outburst in ASASSN-16dt.}
\label{log}
\begin{center}
\begin{tabular}{rrrrrc}
\hline
${\rm Start}^{*}$ & ${\rm End}^{*}$ & ${\rm Mag}^{\dagger}$ & ${\rm Error}^{\ddagger}$ & $N^{\S}$ & ${\rm Obs}^{\parallel}$ \\ \hline
57482.1006 & 57482.2068 & 13.199 & 0.019 & 84 & SPE \\ 
  57483.5701 & 57483.7663 & 13.417 & 0.020 & 110 & HMB \\ 
  57484.5278 & 57484.7410 & 13.501 & 0.029 & 124 & HMB \\ 
  57485.0301 & 57485.2374 & 13.491 & 0.025 & 116 & SPE \\ 
  57485.2343 & 57485.4249 & -0.018 & 0.018 & 549 & MLF \\ 
  57485.5292 & 57485.7608 & 13.563 & 0.021 & 102 & HaC \\ 
  57486.2857 & 57486.4086 & 0.064 & 0.014 & 351 & MLF \\ 
  57486.5348 & 57486.7581 & 13.623 & 0.015 & 106 & HMB \\ 
  57487.2795 & 57487.3615 & 0.122 & 0.012 & 237 & MLF \\ 
  57487.5282 & 57487.7552 & 13.710 & 0.020 & 108 & HaC \\ 
  57488.5268 & 57488.7504 & 13.793 & 0.016 & 106 & HaC \\ 
  57489.2968 & 57489.4274 & 0.282 & 0.014 & 376 & MLF \\ 
  57489.5268 & 57489.7497 & 13.868 & 0.017 & 106 & HaC \\ 
  57490.0690 & 57490.1896 & 13.874 & 0.012 & 94 & SPE \\ 
  57490.5248 & 57490.7470 & 13.925 & 0.019 & 96 & HMB \\ 
  57491.7255 & 57491.7457 & 14.000 & 0.017 & 11 & HaC \\ 
  57492.5232 & 57492.7410 & 14.046 & 0.024 & 95 & HaC \\ 
  57493.2464 & 57493.4544 & 0.561 & 0.018 & 598 & MLF \\ 
  57493.5602 & 57493.7401 & 14.145 & 0.019 & 64 & HaC \\ 
  57494.5236 & 57494.7371 & 14.204 & 0.015 & 137 & HaC \\ 
  57495.5177 & 57495.7343 & 14.262 & 0.032 & 120 & HMB \\ 
  57495.7321 & 57495.8591 & 13.944 & 0.030 & 180 & SGE \\ 
  57496.8774 & 57497.0056 & 14.283 & 0.010 & 162 & MGW \\ 
  57499.5158 & 57499.7530 & 16.774 & 0.352 & 124 & HMB \\ 
  57500.5370 & 57500.7504 & 16.749 & 0.701 & 37 & HaC \\ 
  57500.8773 & 57501.0057 & 18.604 & 0.357 & 156 & MGW \\ 
  57502.8776 & 57503.0052 & 17.742 & 0.129 & 161 & MGW \\ 
  57504.2149 & 57504.4214 & 0.978 & 0.040 & 567 & MLF \\ 
  57504.3673 & 57504.5505 & 14.572 & 0.043 & 207 & deM \\ 
  57504.8605 & 57505.1721 & 14.203 & 0.033 & 389 & MGW \\ 
  57504.9694 & 57505.0386 & -0.567 & 0.054 & 195 & KU1 \\ 
  57505.2143 & 57505.5014 & 0.639 & 0.039 & 823 & MLF \\ 
  57505.2921 & 57505.4930 & 14.223 & 0.046 & 193 & BSM \\ 
  57505.3857 & 57505.5202 & 14.266 & 0.037 & 150 & deM \\ 
  57505.5493 & 57505.7691 & 14.150 & 0.040 & 294 & DKS \\ 
  57505.9719 & 57506.1719 & 14.111 & 0.039 & 248 & MGW \\ 
  57506.5123 & 57506.7358 & 14.223 & 0.036 & 130 & HMB \\ 
  57506.8985 & 57507.1090 & 14.092 & 0.034 & 263 & MGW \\ 
  57507.5114 & 57507.7671 & 14.230 & 0.036 & 103 & HaC \\ 
  57508.5115 & 57508.7649 & 14.285 & 0.036 & 121 & HMB \\ 
  57508.9882 & 57509.1964 & 14.282 & 0.168 & 305 & Ioh \\ 
  57509.5110 & 57509.7620 & 14.341 & 0.033 & 119 & HaC \\ 
  57510.2135 & 57510.4897 & 0.769 & 0.037 & 784 & MLF \\ 
  57510.5097 & 57510.7592 & 14.418 & 0.029 & 136 & HaC \\ 
  57511.0128 & 57511.0548 & -0.425 & 0.045 & 118 & KU1 \\ 
  57511.2339 & 57511.4076 & 0.894 & 0.030 & 497 & MLF \\ 
  57511.5121 & 57511.7565 & 14.487 & 0.028 & 132 & HMB \\ 
  57512.0048 & 57512.0523 & 14.489 & 0.024 & 30 & SPE \\ 
  57512.2163 & 57512.4337 & 0.897 & 0.027 & 619 & MLF \\ 
  \hline
\end{tabular}
\end{center}
\end{table*}

\setcounter{table}{1}
\begin{table*}[htb]
\caption{Log of observations of the 2016 outburst in ASASSN-16dt (continued).}
\label{log}
\begin{center}
\begin{tabular}{rrrrrc}
\hline
${\rm Start}^{*}$ & ${\rm End}^{*}$ & ${\rm Mag}^{\dagger}$ & ${\rm Error}^{\ddagger}$ & $N^{\S}$ & ${\rm Obs}^{\parallel}$ \\ \hline
  57512.5091 & 57512.7530 & 14.549 & 0.029 & 125 & HaC \\ 
  57512.8567 & 57513.1718 & 14.462 & 0.028 & 387 & MGW \\ 
  57512.9488 & 57513.1390 & -0.302 & 0.045 & 522 & KU1 \\ 
  57513.1065 & 57513.1898 & 14.591 & 0.025 & 64 & SPE \\ 
  57513.2467 & 57513.4953 & 1.045 & 0.032 & 717 & MLF \\ 
  57513.3047 & 57513.4691 & 14.573 & 0.041 & 166 & BSM \\ 
  57513.5085 & 57513.7281 & 14.622 & 0.069 & 103 & HaC \\ 
  57513.6198 & 57513.7287 & 14.668 & 0.021 & 85 & BJA \\ 
  57513.8567 & 57514.1253 & 14.539 & 0.021 & 336 & MGW \\ 
  57514.2284 & 57514.4699 & 1.118 & 0.036 & 587 & MLF \\ 
  57514.5454 & 57514.7371 & 14.595 & 0.031 & 212 & DKS \\ 
  57514.5834 & 57514.7549 & 14.493 & 0.040 & 170 & UJH \\ 
  57514.8542 & 57515.1715 & 14.629 & 0.027 & 397 & MGW \\ 
  57515.5091 & 57515.7304 & 14.798 & 0.062 & 113 & HaC \\ 
  57515.5454 & 57515.7477 & 14.692 & 0.037 & 272 & DKS \\ 
  57515.5852 & 57515.6621 & 14.534 & 0.046 & 35 & UJH \\ 
  57516.0414 & 57516.1724 & 14.740 & 0.087 & 45 & Ioh \\ 
  57516.5506 & 57516.7373 & 14.896 & 0.026 & 78 & HaC \\ 
  57516.5512 & 57516.7362 & 14.781 & 0.040 & 187 & DKS \\ 
  57518.5068 & 57518.7364 & 16.521 & 0.168 & 117 & HaC \\ 
  57519.8462 & 57520.0213 & 18.005 & 0.134 & 177 & MGW \\ 
  57520.0184 & 57520.0209 & 3.267 & 0.278 & 3 & KU1 \\ 
  57520.5076 & 57520.7301 & 18.318 & 0.384 & 100 & HaC \\ 
  57521.5070 & 57521.7289 & 18.539 & 0.376 & 98 & HaC \\ 
  57521.8505 & 57522.0214 & 18.451 & 0.152 & 213 & MGW \\ 
  57522.8520 & 57523.1295 & 18.614 & 0.113 & 186 & MGW \\ 
  57530.5033 & 57530.5044 & 18.526 & 0.002 & 2 & HaC \\ 
  57531.5030 & 57531.5041 & 18.530 & 0.040 & 2 & HMB \\ 
  57531.8460 & 57532.1077 & 18.158 & 0.222 & 172 & MGW \\ 
  57532.5027 & 57532.5038 & 15.157 & 0.003 & 2 & HMB \\ 
  57532.9298 & 57533.1074 & 14.994 & 0.061 & 119 & MGW \\ 
  57533.5023 & 57533.5034 & 15.437 & 0.019 & 2 & HaC \\ 
  57534.5026 & 57534.5037 & 16.479 & 0.046 & 2 & HaC \\ 
  57535.5020 & 57535.5031 & 18.445 & 0.302 & 2 & HaC \\ 
  57537.8864 & 57537.8879 & 18.433 & 0.115 & 2 & MGW \\ 
  57538.8439 & 57538.8455 & 18.442 & 0.188 & 2 & MGW \\ 
  57539.8442 & 57539.8457 & 18.512 & 0.283 & 2 & MGW \\ 
  57542.4941 & 57542.4951 & 19.788 & 0.152 & 2 & HaC \\ 
  57543.4998 & 57543.5009 & 19.343 & 0.325 & 2 & HaC \\ 
  57544.4993 & 57544.5004 & 19.203 & 0.249 & 2 & HaC \\ 
  57547.4998 & 57547.5009 & 19.261 & 0.115 & 2 & HaC \\ 
  57548.5008 & 57548.5019 & 19.607 & 0.607 & 2 & HaC \\ 
  57551.4941 & 57551.4952 & 19.230 & 0.112 & 2 & HaC \\ 
\hline
\multicolumn{6}{l}{$^{*}{\rm BJD}-2400000.0$.}\\
\multicolumn{6}{l}{$^{\dagger}$Mean magnitude.}\\
\multicolumn{6}{l}{$^{\ddagger}1\sigma$ of mean magnitude.}\\
\multicolumn{6}{l}{$^{\S}$Number of observations.}\\
\multicolumn{6}{l}{\parbox{275pt}{$^{\parallel}$Observer's code: SPE (Peter Starr), HaC (Franz-Josef Hambsch), MLF (Berto Monard), SGE (Geoff Stone), MGW (Gordon Myers), deM (Enrique de Miguel), KU1 (Kyoto Univ.~Team), BSM (Stephen M.~Brincat), DKS (Shawn Dvorak), Ioh (Hiroshi Itoh), BJA (Boardman James) and UJH (Joseph Ulowetz)}}\\
\end{tabular}
\end{center}
\end{table*}

\begin{table*}[htb]
\caption{Log of observations of the 2016 outburst in ASASSN-16hg.}
\label{log16hg}
\begin{center}
\begin{tabular}{rrrrrc}
  \hline
${\rm Start}^{*}$ & ${\rm End}^{*}$ & ${\rm Mag}^{\dagger}$ & ${\rm Error}^{\ddagger}$ & $N^{\S}$ & ${\rm Obs}^{\parallel}$ \\ \hline
  57482.1006 & 57586.9298 & 14.450 & 0.026 & 126 & HaC \\ 
  57587.5914 & 57587.9316 & 14.573 & 0.033 & 124 & HaC \\ 
  57588.1796 & 57588.3012 & 15.827 & 0.042 & 65 & KU1 \\ 
  57588.5700 & 57588.9314 & 14.673 & 0.042 & 147 & HaC \\ 
  57589.1455 & 57589.2971 & 15.919 & 0.055 & 86 & KU1 \\ 
  57589.6285 & 57589.9299 & 14.838 & 0.038 & 106 & HaC \\ 
  57590.6264 & 57590.9293 & 15.367 & 0.083 & 101 & HaC \\ 
  57591.6232 & 57591.9304 & 14.866 & 0.081 & 118 & HaC \\ 
  57592.6389 & 57592.9300 & 14.733 & 0.050 & 112 & HaC \\ 
  57593.6371 & 57593.9296 & 14.839 & 0.056 & 113 & HaC \\ 
  57594.7362 & 57594.9297 & 14.967 & 0.041 & 65 & HaC \\ 
  57595.6319 & 57595.9309 & 15.075 & 0.047 & 95 & HaC \\ 
  57596.6855 & 57596.9303 & 15.185 & 0.031 & 121 & HaC \\ 
  57597.6264 & 57597.9300 & 15.324 & 0.037 & 137 & HaC \\ 
  57598.6237 & 57598.9295 & 16.262 & 0.144 & 138 & HaC \\ 
  57599.6210 & 57599.9281 & 18.176 & 0.392 & 126 & HaC \\ 
  57600.7491 & 57600.9295 & 18.736 & 0.441 & 89 & HaC \\ 
  57603.7192 & 57603.7192 & 18.403 & -- & 1 & HaC \\ 
  57604.7336 & 57604.7336 & 15.578 & -- & 1 & HaC \\ 
  57605.7310 & 57605.7310 & 16.637 & -- & 1 & HaC \\ 
  57606.7283 & 57606.7283 & 18.031 & -- & 1 & HaC \\ 
  57608.7227 & 57608.7227 & 15.622 & -- & 1 & HaC \\ 
  57609.7199 & 57609.7209 & 16.703 & 0.012 & 2 & HaC \\ 
  57610.7175 & 57610.9250 & 18.200 & 0.250 & 189 & HaC \\ 
   \hline
\multicolumn{6}{l}{$^{*}{\rm BJD}-2400000.0$.}\\
\multicolumn{6}{l}{$^{\dagger}$Mean magnitude.}\\
\multicolumn{6}{l}{$^{\ddagger}1\sigma$ of mean magnitude.}\\
\multicolumn{6}{l}{$^{\S}$Number of observations.}\\
\multicolumn{6}{l}{$^{\parallel}$Observer's code: see the annotation in Table E2.  }\\
\end{tabular}
\end{center}
\end{table*}

\begin{table*}[htb]
\caption{Times of superhump maxima in ASASSN-16dt.  }
\label{max}
\begin{center}
\begin{tabular}{rllrr}
\hline
$E$ & ${\rm Max}^{\dagger}$ & Error & ${O - C}^{\ddagger}$ & $N^{\S}$  \\ \hline
  0 & 57502.9004 & 0.0012 & -0.0247 & 65 \\ 
  1 & 57502.9716 & 0.0008 & -0.0181 & 64 \\ 
  21 & 57504.2745 & 0.0005 & -0.0070 & 149 \\ 
  22 & 57504.3356 & 0.0007 & -0.0104 & 148 \\ 
  23 & 57504.4028 & 0.0007 & -0.0079 & 156 \\ 
  25 & 57504.5329 & 0.0007 & -0.0069 & 50 \\ 
  31 & 57504.9243 & 0.0002 & -0.0031 & 65 \\ 
  32 & 57504.9887 & 0.0002 & -0.0033 & 63 \\ 
  33 & 57505.0547 & 0.0002 & -0.0019 & 65 \\ 
  34 & 57505.1197 & 0.0002 & -0.0015 & 65 \\ 
  36 & 57505.2494 & 0.0004 & -0.0010 & 148 \\ 
  37 & 57505.3147 & 0.0004 & -0.0002 & 200 \\ 
  38 & 57505.3793 & 0.0003 & -0.0002 & 215 \\ 
  39 & 57505.4454 & 0.0003 & 0.0013 & 262 \\ 
  40 & 57505.5101 & 0.0004 & 0.0013 & 108 \\ 
  41 & 57505.5745 & 0.0004 & 0.0011 & 68 \\ 
  42 & 57505.6390 & 0.0003 & 0.0011 & 69 \\ 
  43 & 57505.7049 & 0.0003 & 0.0024 & 70 \\ 
  48 & 57506.0299 & 0.0003 & 0.0045 & 63 \\ 
  49 & 57506.0938 & 0.0006 & 0.0038 & 65 \\ 
  50 & 57506.1587 & 0.0004 & 0.0041 & 47 \\ 
  56 & 57506.5464 & 0.0006 & 0.0042 & 31 \\ 
  57 & 57506.6115 & 0.0005 & 0.0047 & 29 \\ 
  58 & 57506.6779 & 0.0009 & 0.0065 & 29 \\ 
  62 & 57506.9337 & 0.0003 & 0.0039 & 65 \\ 
  63 & 57506.9993 & 0.0003 & 0.0049 & 63 \\ 
  64 & 57507.0630 & 0.0003 & 0.0041 & 64 \\ 
  71 & 57507.5157 & 0.0008 & 0.0046 & 19 \\ 
  72 & 57507.5827 & 0.0005 & 0.0070 & 30 \\ 
  73 & 57507.6449 & 0.0005 & 0.0046 & 14 \\ 
  74 & 57507.7117 & 0.0009 & 0.0068 & 14 \\ 
  87 & 57508.5507 & 0.0005 & 0.0061 & 31 \\ 
  88 & 57508.6136 & 0.0006 & 0.0044 & 22 \\ 
  89 & 57508.6807 & 0.0007 & 0.0069 & 21 \\ 
  90 & 57508.7405 & 0.0010 & 0.0022 & 21 \\ 
  94 & 57509.0044 & 0.0010 & 0.0077 & 60 \\ 
  95 & 57509.0661 & 0.0007 & 0.0048 & 90 \\ 
  96 & 57509.1266 & 0.0007 & 0.0007 & 95 \\ 
  102 & 57509.5199 & 0.0015 & 0.0064 & 23 \\ 
  103 & 57509.5855 & 0.0007 & 0.0075 & 28 \\ 
  104 & 57509.6483 & 0.0012 & 0.0057 & 20 \\ 
  105 & 57509.7150 & 0.0010 & 0.0078 & 21 \\ 
  113 & 57510.2308 & 0.0003 & 0.0069 & 127 \\ 
  114 & 57510.2936 & 0.0003 & 0.0050 & 149 \\ 
  116 & 57510.4223 & 0.0003 & 0.0045 & 148 \\ 
  118 & 57510.5519 & 0.0005 & 0.0050 & 42 \\ 
  119 & 57510.6181 & 0.0009 & 0.0066 & 21 \\ 
  120 & 57510.6829 & 0.0008 & 0.0068 & 21 \\ 
  121 & 57510.7510 & 0.0016 & 0.0102 & 17 \\ 
  129 & 57511.2585 & 0.0005 & 0.0010 & 145 \\ 
  130 & 57511.3257 & 0.0005 & 0.0036 & 149 \\ 
\hline
\end{tabular}
\end{center}
\end{table*}

\setcounter{table}{3}
\begin{table*}[htb]
\caption{Times of superhump maxima in ASASSN-16dt (continued).  }
\label{max}
\begin{center}
\begin{tabular}{rllrr}
\hline
$E$ & ${\rm Max}^{\dagger}$ & Error & ${O - C}^{\ddagger}$ & $N^{\S}$  \\ \hline
  131 & 57511.3881 & 0.0006 & 0.0015 & 128 \\ 
  134 & 57511.5819 & 0.0007 & 0.0015 & 37 \\ 
  135 & 57511.6469 & 0.0016 & 0.0019 & 21 \\ 
  136 & 57511.7104 & 0.0013 & 0.0008 & 21 \\ 
  141 & 57512.0316 & 0.0013 & -0.0010 & 30 \\ 
  144 & 57512.2265 & 0.0009 & 0.0002 & 120 \\ 
  145 & 57512.2919 & 0.0004 & 0.0010 & 147 \\ 
  146 & 57512.3573 & 0.0004 & 0.0017 & 149 \\ 
  147 & 57512.4256 & 0.0006 & 0.0055 & 106 \\ 
  149 & 57512.5517 & 0.0006 & 0.0024 & 42 \\ 
  150 & 57512.6152 & 0.0022 & 0.0013 & 19 \\ 
  151 & 57512.6799 & 0.0010 & 0.0014 & 19 \\ 
  152 & 57512.7418 & 0.0020 & -0.0013 & 15 \\ 
  154 & 57512.8690 & 0.0003 & -0.0033 & 48 \\ 
  155 & 57512.9360 & 0.0004 & -0.0009 & 65 \\ 
  156 & 57513.0018 & 0.0004 & 0.0003 & 62 \\ 
  157 & 57513.0665 & 0.0005 & 0.0004 & 65 \\ 
  158 & 57513.1325 & 0.0004 & 0.0019 & 103 \\ 
  163 & 57513.4509 & 0.0005 & -0.0027 & 199 \\ 
  166 & 57513.6463 & 0.0006 & -0.0011 & 55 \\ 
  167 & 57513.7092 & 0.0007 & -0.0028 & 38 \\ 
  170 & 57513.9048 & 0.0004 & -0.0010 & 65 \\ 
  171 & 57513.9709 & 0.0004 & 0.0005 & 63 \\ 
  172 & 57514.0354 & 0.0005 & 0.0005 & 65 \\ 
  173 & 57514.0967 & 0.0005 & -0.0028 & 65 \\ 
  175 & 57514.2292 & 0.0040 & 0.0005 & 69 \\ 
  176 & 57514.2910 & 0.0008 & -0.0023 & 123 \\ 
  177 & 57514.3561 & 0.0013 & -0.0018 & 115 \\ 
  178 & 57514.4225 & 0.0008 & 0.0000 & 149 \\ 
  181 & 57514.6159 & 0.0010 & -0.0004 & 176 \\ 
  182 & 57514.6785 & 0.0012 & -0.0023 & 136 \\ 
  183 & 57514.7412 & 0.0012 & -0.0043 & 116 \\ 
  185 & 57514.8640 & 0.0011 & -0.0107 & 47 \\ 
  186 & 57514.9381 & 0.0006 & -0.0011 & 65 \\ 
  187 & 57515.0012 & 0.0004 & -0.0026 & 63 \\ 
  188 & 57515.0648 & 0.0006 & -0.0036 & 65 \\ 
  189 & 57515.1308 & 0.0006 & -0.0022 & 65 \\ 
  196 & 57515.5853 & 0.0025 & 0.0001 & 117 \\ 
  197 & 57515.6400 & 0.0012 & -0.0098 & 97 \\ 
  198 & 57515.7086 & 0.0015 & -0.0057 & 88 \\ 
  211 & 57516.5537 & 0.0020 & -0.0003 & 53 \\ 
  212 & 57516.6127 & 0.0025 & -0.0059 & 122 \\ 
  213 & 57516.6737 & 0.0011 & -0.0095 & 124 \\ 
  214 & 57516.7433 & 0.0043 & -0.0045 & 92 \\ 
  263 & 57519.9120 & 0.0008 & -0.0008 & 64 \\ 
  264 & 57519.9789 & 0.0021 & 0.0015 & 44 \\ 
  293 & 57521.8583 & 0.0007 & 0.0077 & 44 \\ 
  294 & 57521.9259 & 0.0014 & 0.0107 & 65 \\ 
  295 & 57521.9903 & 0.0009 & 0.0105 & 64 \\ 
\hline
\multicolumn{5}{l}{$^{*}$Cycle counts.}\\
\multicolumn{5}{l}{$^{\dagger}$BJD$-$2400000.0.}\\
\multicolumn{5}{l}{$^{\ddagger}$ C = 2457502.925071$+$0.0653055 E.}\\
\multicolumn{5}{l}{\parbox{170pt}{$^{\S}$Number of points used for determining the maximum.}}\\
\end{tabular}
\end{center}
\end{table*}

\begin{table*}[b]
\caption{Times of superhump maxima in ASASSN-16hg.  }
\label{max16hg}
\begin{center}
\begin{tabular}{rllrr}
\hline
$E$ & ${\rm Max}^{\dagger}$ & Error & ${O - C}^{\ddagger}$ & $N^{\S}$  \\ \hline
  0 & 57591.6600 & 0.0065 & -0.0010 & 18 \\ 
  1 & 57591.7237 & 0.0037 & 0.0003 & 18 \\ 
  2 & 57591.7863 & 0.0013 & 0.0006 & 19 \\ 
  3 & 57591.8442 & 0.0015 & -0.0038 & 21 \\ 
  4 & 57591.9102 & 0.0015 & -0.0002 & 20 \\ 
  16 & 57592.6570 & 0.0012 & -0.0015 & 16 \\ 
  17 & 57592.7208 & 0.0014 & -0.0001 & 18 \\ 
  18 & 57592.7796 & 0.0016 & -0.0037 & 19 \\ 
  19 & 57592.8439 & 0.0017 & -0.0017 & 21 \\ 
  20 & 57592.9060 & 0.0016 & -0.0020 & 21 \\ 
  32 & 57593.6548 & 0.0022 & -0.0013 & 15 \\ 
  33 & 57593.7179 & 0.0011 & -0.0006 & 18 \\ 
  34 & 57593.7767 & 0.0012 & -0.0041 & 19 \\ 
  35 & 57593.8421 & 0.0017 & -0.0011 & 20 \\ 
  36 & 57593.9059 & 0.0016 & 0.0004 & 20 \\ 
  50 & 57594.7769 & 0.0021 & -0.0015 & 16 \\ 
  51 & 57594.8361 & 0.0018 & -0.0046 & 17 \\ 
  52 & 57594.9042 & 0.0015 & 0.0011 & 17 \\ 
  64 & 57595.6516 & 0.0017 & 0.0003 & 14 \\ 
  65 & 57595.7146 & 0.0048 & 0.0010 & 15 \\ 
  66 & 57595.7729 & 0.0016 & -0.0030 & 16 \\ 
  67 & 57595.8356 & 0.0015 & -0.0027 & 16 \\ 
  68 & 57595.9068 & 0.0015 & 0.0061 & 15 \\ 
  81 & 57596.7117 & 0.0018 & 0.0006 & 15 \\ 
  82 & 57596.7738 & 0.0032 & 0.0003 & 27 \\ 
  83 & 57596.8389 & 0.0030 & 0.0030 & 28 \\ 
  84 & 57596.8956 & 0.0022 & -0.0026 & 28 \\ 
  96 & 57597.6508 & 0.0027 & 0.0044 & 14 \\ 
  97 & 57597.7086 & 0.0029 & -0.0001 & 15 \\ 
  98 & 57597.7678 & 0.0033 & -0.0032 & 25 \\ 
  99 & 57597.8382 & 0.0059 & 0.0048 & 28 \\ 
  100 & 57597.8903 & 0.0047 & -0.0055 & 28 \\ 
  114 & 57598.7596 & 0.0037 & -0.0090 & 23 \\ 
  \hline
\multicolumn{5}{l}{$^{*}$Cycle counts.}\\
\multicolumn{5}{l}{$^{\dagger}$BJD$-$2400000.0.}\\
\multicolumn{5}{l}{$^{\ddagger}$ C = 2457591.6610$+$0.0623475 E.}\\
\multicolumn{5}{l}{$^{\S}$\parbox{170pt}{Number of points used for determining the maximum.}}\\
\end{tabular}
\end{center}
\end{table*}

\end{document}